**General rights:**

# From Pre-Quantum to Post-Quantum IoT Security: A Survey on Quantum-Resistant Cryptosystems for the Internet of Things

Tiago M. Fernández-Caramés, *Senior Member, IEEE*

*Abstract*—Although quantum computing is still in its nascent age, its evolution threatens the most popular public-key encryption systems. Such systems are essential for today's Internet security due to their ability for solving the key distribution problem and for providing high security in non-secure communications channels that allow for accessing websites or for exchanging e-mails, financial transactions, digitally-signed documents, military communications or medical data. Cryptosystems like RSA (Rivest-Shamir-Adleman), ECC (Elliptic Curve Cryptography) or Diffie-Hellman have spread worldwide and are part of diverse key Internet standards like Transport Layer Security (TLS), which are used both by traditional computers and IoT devices. It is especially difficult to provide high security to IoT devices, mainly because many of them rely on batteries and are resource-constrained in terms of computational power and memory, what implies that specific energy-efficient and lightweight algorithms need to be designed and implemented for them. These restrictions become relevant challenges when implementing cryptosystems that involve intensive mathematical operations and demand substantial computational resources, which are often required in applications where data privacy has to be preserved for the long term, like IoT applications for Defense, mission-critical scenarios or smart healthcare. Quantum computing threatens such a long-term IoT device security and researchers are currently developing solutions to mitigate such a threat. This article provides a survey on what can be called post-quantum IoT systems (IoT systems protected from the currently known quantum computing attacks): the main post-quantum cryptosystems and initiatives are reviewed, the most relevant IoT architectures and challenges are analyzed, and the expected future trends are indicated. Thus, this paper is aimed at providing a wide view of post-quantum IoT security and give useful guidelines to the future post-quantum IoT developers.

## I. INTRODUCTION

The popularity of Internet of Things (IoT) is growing so fast that some reports estimate that 75 billion IoT devices will be in operation in 2025 [1]. This enormous growth will require to standardize protocols and develop the appropriate architectures to provide services to IoT devices. In general, such devices can be defined as battery dependent and mostly resource-constrained in terms of computational power and memory, so

Tiago M. Fernández-Caramés is with the Department of Computer Engineering, Faculty of Computer Science, Centro de Investigación CITIC, Universidade da Coruña, 15071, A Coruña, Spain. (e-mail: tiago.fernandez@udc.es).

This work has been funded by the Agencia Estatal de Investigación of Spain (TEC2016-75067-C4-1-R) and ERDF funds of the EU (AEI/FEDER, UE).
Manuscript received XXX; revised XXX.

their most complex processing tasks are usually carried out in centralized servers or clouds on the Internet.

To preserve IoT node security, hash functions, symmetric cryptography and public-key cryptosystems (i.e., asymmetric cryptographic systems) are typically used. In the case of public-key cryptography, since the first practical encryption systems were made public during the 70s [2], [3], public-key cryptography has become essential for the current Internet communications due to its ability to provide high security to websites [4], e-mails [5], financial transactions [6], digitally-signed documents [7], military communications [8] or medical data [9]. Thus, public-key cryptosystems like RSA (Rivest-Shamir-Adleman) [2], ECC (Elliptic Curve Cryptography) [10], [11] or Diffie-Hellman (DH) [3] have spread worldwide and are part of diverse key Internet standards like Transport Layer Security (TLS) [4], which are used both by traditional computers and IoT devices.

However, recent advances in computing and communications have made it easier to reach the computational effort required to break certain asymmetric schemes, what derived into increasing the recommended minimum key size. For instance, the minimum recommended RSA key size is currently between 2048 and 4096 bits (depending on the sort of information to be protected), since 768-bit and 1024-bit RSA implementations were broken around 2010 [12], [13]. Key-size growth is a solution for the short-term, until technology catches up and provides the required computational effort. The problem is that, for some countries and entities, certain critical information (e.g., national security information) needs to remain secret for long periods of time [14], what requires cryptographic systems that guarantee that data will remain secret for the long term. That is the reason indicated in July 2015 by the National Security Agency (NSA) [15], which, after analyzing the impact of quantum computing on information assurance (IA) and IA-enabled IT products, recommended increasing the ECC security level to be used in their Suite B set of cryptographic algorithms while transitioning to quantum-resistant alternatives. Although there has been some speculation on NSA's announcement [16] and on the fact that Edward Snowden's leaks showed no critical advances of the NSA on quantum computing [17], it is estimated that, in the next 20 years, quantum computers will be enough functional to break current strong public-key cryptosystems easily [18].

Due to the previously mentioned issues, it was coined the term post-quantum cryptography (also called quantum-proof, quantum-safe or quantum-resistant cryptography), which refers to cryptographic algorithms that are said to be secure when attacked by a quantum computer.

This article reviews the state of the art on post-quantum cryptography for IoT systems. There are other reviews and surveys on post-quantum computing in the literature [19], [20], [21], [22], but, although they provide useful information, they are not IoT-specific [19], are introductory [20], [21] or are focused on a type of post-quantum cryptosystem [19], [22].

In contrast, this article presents a thorough revision on the most relevant aspects that influence post-quantum IoT developments and provides together the following main contributions, which have not been found together in the previous literature:

- After detailing the basics on the impact of post-quantum cryptography on IoT, a survey on the main post-quantum initiatives is provided.
- An analysis is carried out on the latest IoT communications architectures in terms of the links that need to be secured in a post-quantum scenario.
- The main types of post-quantum public-key cryptosystems are described and the main advantages and disadvantages of their application to resource-constrained IoT devices are studied.
- Thorough analyses and comparisons are provided on the performance of the NIST second-round post-quantum candidates when implementing them on low-power, medium-power and high-power IoT devices.
- Multiple tables are included throughout the paper in order to simplify the selection of the most appropriate post-quantum schemes according to the computational resources available in an IoT device.
- An detailed study on the main challenges and future trends for post-quantum IoT cryptosystem development is provided.

The rest of this paper is structured as follows. Section II analyzes the need for moving from pre-quantum to post-quantum IoT systems. Section III reviews the most relevant post-quantum cryptography projects and standardization initiatives. Section IV provides a thorough review on the main types of post-quantum cryptosystems and compares the most relevant implementations for different hardware platforms. Section V studies the most common and some of the latest IoT architectures in order to emphasize the need for securing them in a post-quantum world. Section VI analyzes the state of the art of previous IoT implementations and compares their performance. Section VII enumerates the main post-quantum IoT challenges and promising future trends. Finally, Section VIII is devoted to conclusions.

## II. PRE-QUANTUM VERSUS POST-QUANTUM IoT SECURITY

IoT security has to deal with different types of attacks [23], [24], [25], but nowadays, in terms of communications, it relies essentially on asymmetric and symmetric cryptosystems. The strength of such cryptosystems has been traditionally associated with their bits-of-security level, which is a measure of the computational strength required to break a cryptosystem by using brute force through classical computers (e.g., a cryptosystem is said to have a 256-bit security if the difficulty of attacking it with a classical computer is similar (in terms of time and computational resources) to perform a brute-force search attack on a 256-bit cryptographic key). For instance, Table I compares the security levels of popular symmetric (TDEA (Triple Data Encryption Algorithm), AES (Advanced Encryption Standard)) and asymmetric cryptosystems (RSA, ECDSA (Elliptic Curve Digital Signature Algorithm)) for the same bits-of-security level.

Pre-quantum symmetric algorithms and hash functions seem to be still valid for the post-quantum era: it is considered unlikely that efficient quantum algorithms will be found for NP-hard problems [26]. In fact, it is assumed that symmetric algorithms and hash functions will only require to increase their key size/output [14]. For instance, in [27] the authors describe a quantum birthday attack that creates a $\sqrt[3]{N}$ size table and uses Grover's algorithm [28] to find collisions in hash functions: the authors conclude that such functions would need to generate a 3*n-bit output to provide a n-bit security level, what disqualifies as quantum-safe many current hash functions (however, relevant hash functions like SHA-2 and SHA-3 will remain quantum resistant by increasing their outputs).

In the case of public-key cryptography, it makes use of pairs of keys: a public key is used for encrypting messages addressed to a specific user and a private key is used by such a user to decrypt them. Since a public key and a private key are related mathematically, the strength of a public-key cryptosystem depends on the computational effort (called "work factor" in cryptography) required to perform a brute-force key search attack to find a private key from its paired public key. Therefore, public-key cryptography relies on mathematical problems like integer factorization, discrete logarithms or elliptic curves, which, until recently, had no efficient solution.

Public-key cryptosystems main advantage in insecure networks is that they solve the key distribution problem due to being asymmetric: the public key does not allow for decrypting messages. In contrast, symmetric cryptography uses the same key for encrypting and decrypting messages, hence it requires secure ways to store and deliver keys among the peers interested in exchanging information. However, the keys of a public-key cryptosystem need to have a special structure (e.g., large primes), so its generation is usually far more expensive than in the case of symmetric cryptosystems, which often use randomly generated k-bit strings as keys.

Nonetheless, quantum computing threatens the most popular public-key encryption systems [15]. Quantum attacks affect the most popular public-key algorithms, including RSA, ECDSA, ECDH (Elliptic Curve Diffie-Hellman), DSA (Digital Signature Algorithm) [29] and others based on the integer factorization problem, the elliptic-curve discrete logarithm problem or the discrete logarithm problem. All of them can be solved fast with Shor's algorithm [30] on a sufficiently powerful quantum computer. In addition, quantum computers can make use of Grover's algorithm [28] to speed up brute force attacks on symmetric ciphers by roughly a quadratic factor [18].

TABLE I
COMPARABLE SYMMETRIC AND ASYMMETRIC CRYPTOSYSTEMS DEPENDING OF THEIR SECURITY LEVEL (SOURCE: [32]).

| Security Level | Symmetric Cryptosystem (Key Size) | RSA (Key Size) | ECDSA Curve (Key Size) |
|---|---|---|---|
| 80 | 2TDEA (112 bits) | 1,024 bits | prime192v1 (192 bits) |
| 112 | 3TDEA (168 bits) | 2,048 bits | secp224r1 (224 bits) |
| 128 | AES-128 (128 bits) | 3,072 bits | secp256r1 (256 bits) |
| 192 | AES-192 (192 bits) | 7,680 bits | secp384r1 (384 bits) |

In addition, it must be noted that current 80-bit security cryptosystems can be broken at a cost ranging from tens of thousands to hundreds of millions of dollars [14] and it is estimated that 112-bit security systems will remain secure for classical computer attacks for the next 30 to 40 years [14]. Nonetheless, according to [31], 160-bit elliptic curves could be broken on a quantum computer with around 1,000 qubits, while factorizing 1024-bit RSA would require about 2,000 qubits, which are far more than what today's quantum computers provide (as of writing, IonQ claims to have the largest quantum computer, which provides 79 qubits). Therefore, since computationally powerful quantum computers are not expected to be available in the next 20 years [18], it is considered more urgent to transition to post-quantum cryptosystems that also withstand classical computer attacks, than improving traditional cryptosystems. Due to this reason, in the last years some researchers already studied the security threat of quantum computers in different fields [19], [33], [34], [35], [36].

## III. POST-QUANTUM INITIATIVES

Post-quantum computing is currently a promising research area fostered since 2006 by its own conference series (PQCrypto) and by different entities that are carrying out diverse projects and standardization activities. Although most of such activities are not explicitly focused on post-quantum IoT systems, it is worth monitoring their outputs and standardization processes.

### A. Post-quantum cryptography projects

Substantial financial support has been granted by the European Union (EU) and the Japanese Science and Technology Agency to projects related to post-quantum cryptosystems, like PQCrypto [37], SAFEcrypto [38], CryptoMathCREST [39] or PROMETHEUS [40].

In the case of PQCrypto, it is a project funded with roughly € 4 M from March 2015 to February 2018 by the EU through the Horizon 2020 program. The project was coordinated from the Netherlands by Eindhoven University of Technology, but there were participants from Belgium, Germany, Denmark, France, Israel and Taiwan. The project was divided into three work packages aimed at investigating post-quantum cryptography for Internet communications, for cloud computing and for low-power embedded devices. The results of the project include 27 journal articles, 45 papers on conferences, multiple reports and various implementations [41].

SAFECrypto shared similar objectives with PQCrypto (to provide new post-quantum cryptosystems for long-term security), but in this case, the researchers focused on lattice problems as the source of computational hardness. The project was also funded by the EU (with € 3.2 M) and its tasks were carried out from January 2015 to December 2018. SAFECrypto was led by the Queen's University of Belfast (United Kingdom) with partners from Switzerland, France, Germany, United Kingdom and Ireland. The obtained results were published in 7 journal articles and 12 conference papers and a number of reports [42]. The project software library is available in GitHub [43] and includes implementations of Ring-LWE (Ring Learning With Errors), BLISS-B (Bimodal Lattice Signature Scheme B) or Kyber.

Regarding CryptoMathCREST, it is a project funded by the Japan Science and Technology Agency since 2015 and that is participated by the University of Tokyo, Kyushu University and the Tokyo Institute of Technology. The project is focused on the study of the mathematical problems underlying the security modeling of the next-generation of cryptographic systems, including post-quantum cryptosystems. The project is divided into fours working groups for security evaluation, mathematical modeling, security modeling and cryptographic applications. The project seems to be quite active in raising awareness on post-quantum security, by organizing and contributing to a significant number of workshops, conferences and invited talks. Moreover, the project partners have published a number papers, although many of them are only available in Japanese.

Finally, it is worth mentioning the PROMETHEUS project, which started in January 2018 and will continue its activities until December 2021. The project is funded by the EU with roughly € 5.5 M and is coordinated by the École Normale Supérieure de Lyon, which collaborates with relevant public entities and companies from Israel, France, Spain, Germany, United Kingdom or the Netherlands. Although PROMETHEUS started recently, it has already published a relevant number of papers [44].

After analyzing the previously mentioned projects and despite the achieved significant advances, only a small part of the effort of some projects (PQCrypto and SAFECrypto) was put into addressing the specific challenges that arise when making use of post-quantum schemes on resource-constrained devices: the analyzed post-quantum security projects were more focused on computational resource consumption than in other IoT challenges like energy consumption.

### B. Standardization initiatives

*1) ETSI initiatives:* The European Telecommunications Standards Institute (ETSI) released white papers related to quantum security [45], [46]. In addition, together with the Institute for Quantum Computing (IQC), it has organized since 2013 the different editions of the Quantum-Safe Cryptography Workshops [47]. Moreover, ETSI held an Industry Specification Group (ISG) for quantum-safe cryptography [48] until 2017, when its activities were transferred to the ETSI Technical Committee Cyber Working Group on Quantum-Safe Cryptography [49], which has published a number of really interesting deliverables on topics like the limits of quantum computing when applied to symmetric cryptography [50] or on the impact of quantum computing attacks on diverse fields [51].

*2) NIST initiatives:* The U.S. National Institute of Standards and Technology (NIST) has also released reports on the quantum threat [14] and has organized workshops on post-quantum cryprography [52] and its standardization [53]. Moreover, in December 2016 NIST announced a call for proposals for post-quantum public-key cryptosystems [54]. Such a call received 69 candidates in the first round and, as of writing, only 26 of them moved into the second round (17 public-key encryption and key-establishment algorithms, and 9 digital signature schemes) [55]. It is expected that such a standardization process will continue with a third round in 2020/2021 (first drafts are expected to be available between 2022 and 2024).

*3) IETF initiatives:* The Internet Engineering Task Force (IETF) is currently working (together with the Crypto Forum Research Group (CFRG) [56]) on a number of Internet-Drafts on quantum cryptography. For instance, in [57] it is detailed a post-quantum cryptosystem for TLS that combines NTRUEncrypt with RSA/DH. Another example is the Internet-Draft in [58], which proposes to extend the Internet key exchange protocol IKEv2 to be post-quantum. Other Internet-Drafts are focused on topics like the transition from classical to post-quantum cryptography [59] or on implementations such as XMSS (eXtended Merkle Signature Scheme) (RFC 8391) [60] or Leighton-Micali Hash-Based Signatures (RFC 8554) [61].

*4) Other standardization initiatives:* There are other ongoing significant initiatives like the ones fostered by the International Organization for Standardization (ISO) through the ISO/IEC JTC 1/SC 27 (Working group on IT Security techniques) [62] or by the Institute of Electrical and Electronics Engineers (IEEE) through the P1363 project, which has led the release of standards like IEEE 1363.1-2008 for lattice-based public-key cryptography [63]. In addition, the American National Standards Institute (ANSI) Accredited Standards Committee (ASC) X9 released a white paper on the security risks of quantum computing for the financial industry [64] after specifying the use of NTRU for such an industry [65].

There are other relevant ongoing or already finished post-quantum standardization initiatives, most of which have been already reviewed (until April 2018) by the PQCrypto project [66].

## IV. POST-QUANTUM PUBLIC-KEY CRYPTOSYSTEMS

There are currently five main types of post-quantum cryptosystems: code-based, lattice-based, supersingular elliptic curve isogeny, multivariate and hybrid schemes. In the next subsections such schemes are analyzed in relation to their applicability to IoT. As a summary, the most relevant post-quantum types and implementation variants are illustrated in Figure 1.

### A. Post-quantum code-based cryptosystems

Code-based cryptosystems are mainly based on the theory behind error-correction codes, which have provided redundancy to digital communications for a long time. A relevant code-based cryptosystem is McEliece's [67]. Such a cryptosystem is built on binary Goppa codes [68] and its security relies on the syndrome decoding problem (decoding codewords without knowledge of the coding scheme has been proven to be NP-complete [69]). McEliece's cryptosystem is really fast for encryption and reasonably fast for decryption, but it is necessary to tackle its major drawback when implementing it in resource-constrained IoT devices: it makes use of large matrices as public and private keys (their size is usually between 100 kilobytes and several megabytes). To solve this issue, different compression/decompression techniques may be assessed and different versions of the McEliece scheme could be proposed based on other codes (e.g., LDPC (Low-Density Parity-Check) codes, MDPC (Moderate-Density Parity-Check) codes), on the use of quasi-cyclic codes (e.g., QC-LDPC, QC-MDPC or QC-LRPC (Quasi-Cyclic Low-Rank Parity-Check) codes) or by making use of certain coding techniques (e.g., puncturing [70]).

There are also code-based signing algorithms. For instance, the development of variants of the Niederreiter [71] and CFS (Courtois, Finiasz, Sendrier) [72] cryptosystems are specially interesting, since they are very similar to McEliece's scheme. In the case of the CFS variants, it has to be taken into account for IoT developments the fact that, although the generated signatures have a short length and they can be verified really fast, the required key size is very large and signature generation is inefficient. In addition, it should be considered the development of IoT signature schemes derived from the application of the Fiat-Shamir transformation on identification protocols [73], which has already been proved to be useful when creating schemes that outperform CFS [74].

### B. Post-quantum multivariate-based cryptosystems

Multivariate-based cryptosystems are based on the difficulty of solving systems of multivariate equations, which are proven to be NP-hard or NP-complete [75]. It is currently necessary to analyze and implement multivariate encryption and signature schemes that tackle their main drawbacks for IoT applications: their usual decryption inefficiency on resource-constrained devices (which is related to the required decryption "guess work"), their frequent large key size (and, as a consequence, the potential increases in energy consumption) and the existence of large ciphertext overheads.

Some interesting multivariate-based cryptosystems for IoT applications are the ones based on the use of square matrices with random quadratic polynomials, the ones based on the Matsumoto-Imai algorithm and the ones based on

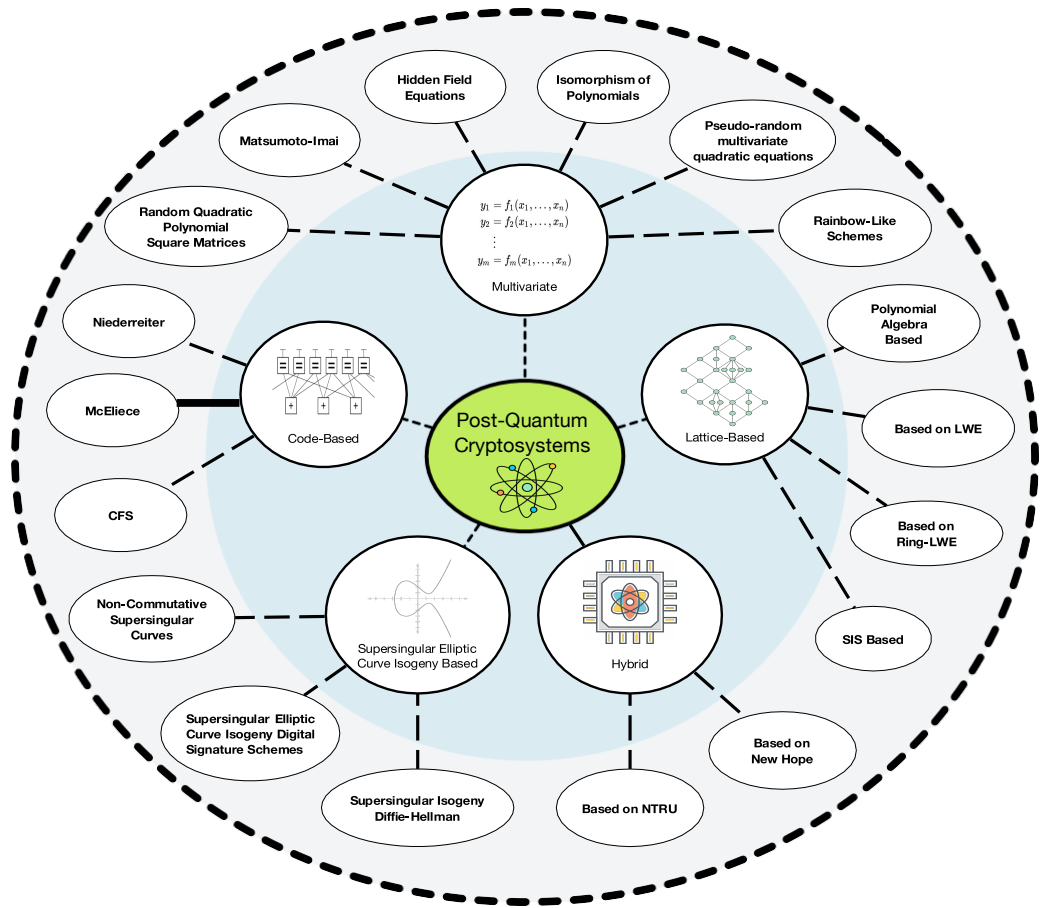

Fig. 1. Most relevant types and implementations of post-quantum public-key cryptosystems and digital signature schemes.

Hidden Field Equations (HFE) [76], [77], [78]. With respect to multivariate digital-signature schemes, there are also variants based on the Matsumoto-Imai algorithm, on HFE and on Isomorphism of Polynomials (IP) [79], which can generate secure signatures with similar sizes to the ones currently based on RSA and ECC. In addition, other cryptosystems should be considered for future IoT developments, like the ones based on pseudo-random multivariate quadratic equations [80] and on Rainbow-Like digital signature schemes, in which successive sets of central variables are obtained from the previous ones by solving linear equations and which have proved to lead to efficient schemes that perform well on resource-constrained systems (e.g., TTS (Tame Transformation Signature) [81], TRMS (Tractable Rational Map Signature) [82] or Rainbow [83]). However, in these latter schemes compression techniques and size optimizations will need to be proposed, since they make use of very large key sizes in comparison to traditional cryptosystem like RSA and ECC (for instance, the Rainbow signature scheme, with n=42, m=24 and q=256, has a public-key size of 22,680 bytes, which is more of what many low-power IoT devices have for their whole flash memory).

### C. Post-quantum lattice-based cryptosystems

Lattice-based cryptosystems are based on lattices, which are sets of points in n-dimensional spaces with a periodic structure. The application of lattices to cryptography is related to the presumed hardness of lattice problems like the Shortest Vector Problem (SVP), the Closest Vector Problem (CVP) or the Shortest Independent Vectors Problem (SIVP) [84]. For instance, in the case of the SVP, the problem consists in finding the shortest non-zero vector within the lattice, which is an NP-hard problem that cannot be currently solved through a quantum algorithm.

Lattice-based cryptosystems present strong security proofs and their implementation is usually simple, fast and relatively efficient. However, future post-quantum lattice-based cryptosystems for IoT devices should manage efficiently the storage and operation with large keys and large ciphertext overheads. For instance, currently, it seems really interesting the study and development of IoT-optimized lattice-based cryptosystems based on polynomial algebra [85], [86], [87], [88], [89] and on the Learning With Errors (LWE) problem and its variants (e.g., LP-LWE (Lindner-Peikert LWE) or Ring-LWE [90], [91]). Compression techniques and optimizations need to be applied when using the previously mentioned cryptosystems in IoT applications, since the key sizes of these lattice-based schemes are lengthier than the typical size of a pre-quantum cryptosystem, although they are clearly smaller than the ones used by code-based or multivariate public-key cryptosystems (for instance, lattice-based cryptosystems like NTRU [85] and NewHope [92] usually make use of a key size in the order of a few thousand bits).

Regarding lattice-based signature schemes, the ones based on Short Integer Solution (SIS) [93] seem to be promising for carrying out signing operations. This is due to the fact that SIS-based solutions allow for creating lattice-based digital signature schemes (based on the Lyubashevsky's signature scheme [94] but introducing new rejection sampling algorithms [95]), which have demonstrated that it is possible to make use of manageable key sizes (roughly 0.6 KB for the public key and 0.25 KB for the private key). Nonetheless, so far, this latter scheme has only been tested on very specific and relatively fast embedded devices (on a Xilinx Spartan-6 FPGA (Field-Programmable Gate Array)) [96], so future developers will have to redesigned it and optimized it for performing fast energy-efficient signing operations in low-power IoT devices. Similarly, lattice-based key-exchange protocols like the ones proposed by Ding et al. [97] with the tweak proposed by Peikert [98] (which has been used by BCNS [99]), would need to be adapted to minimize energy consumption and reduce the required computational resources.

### D. Post-quantum supersingular elliptic curve isogeny cryptosystems

This kind of cryptosystems derive from the isogeny protocol for ordinary elliptic curves proposed in [100]. However, in order to prevent the quantum attack in [101], the supersingular curves need to be non-commutative (the attack relies on the fact that the endomorphism ring is commutative, which is not the case for a supersingular curve whose endomorphism ring is isomorphic to an order in a quaternion algebra), what makes them a promising candidate for implementing post-quantum systems [102], [103]. This kind of cryptosystems are estimated to have a key size in the order of a few thousand bits [104], so, for IoT developments, it will be necessary to study the use of compression techniques and optimizations to reduce key size.

It is also possible to use supersingular elliptic curve isogenies for creating post-quantum digital signature schemes [105], which have to be optimized to be used by resource-constrained devices. Moreover, it is necessary to address the challenges that arise during the implementation of energy-efficient supersingular elliptic curve isogeny cryptosystems and the application of Supersingular Isogeny Diffie-Hellman (SIDH) in resource-constrained IoT devices. Part of such challenges are related to the performance required by key compression schemes for isogeny-based cryptosystems, which may involve computationally intensive steps [106], [107].

### E. Post-quantum hybrid cryptosystems

This kind of cryptographic systems merge pre-quantum and post-quantum cryptosystems in order to provide a double protection to the exchanged data (this is carried out to protect the exchanged data from attacks against post-quantum cryptosystems whose security is still being evaluated). These hybrid systems seem to be the next step prior to full post-quantum security (i.e., security based on sufficiently validated post-quantum algorithms) and they have already been tested by Google [108] or the Tor project [109]. For instance, in the case of Google, an algorithm named CECPQ1 merged New Hope [92], a post-quantum key-exchange algorithm, with X25519, an elliptic curve-based Diffie-Hellman key agreement scheme. Thus, Google guaranteed backward compatibility and integration with TLS at the same time. It is currently being tested CECPQ2, which merges X25519 with instantiations of NTRU (HRSS (Hülsing, Rijneveld, Schanck, Schwabe) and SXY (Saito, Xagawa,Yamakawa)) for the post-quantum part. However, CECPQ1, CECPQ2 and other hybrid cryptosystems were not designed with energy efficiency and IoT resource-constrained devices in mind: they require to implement not only one, but two computationally-intensive cryptosystems that are not necessarily energy efficient. To fill such a gap, it is still necessary to design and to implement hybrid energy-efficient post-quantum IoT cryptosystems that merge the most promising post-quantum and standard pre-quantum public-key schemes.

In this hybrid scenario, it will be especially needed to address the challenge that suppose the exchange of large payloads (due to public-key and ciphertext size) in the context of TLS and IoT architectures, which may derive into dropping messages and potential DoS (Denial-of-Service) attacks.

### F. Cryptosystem comparison

Not all the cryptosystems mentioned in the previous subsections are suitable for IoT systems. Due to this reason, Table II compares some of the most relevant post-quantum cryptosystems proposed so far. Specifically, Table II contains compares implementations of the 17 second-round candidate public-key encryption and key-establishment algorithms that are currently competing in the NIST's post-quantum standardization call. There are many other potential post-quantum cryptosystems (a good compilation and basic description of other relevant cryptosystems can be found in [110]), but it is interesting to focus on the algorithms selected by NIST, since they have already been thoroughly analyzed by the cryptographic community and they have a good chance to become standardized. Table II includes references to the following post-quantum cryptosystems:

- BIKE [111]: is a code-based key encapsulation suite based on QC-MDPC codes.
- Classic McEliece [112]: it is a Key Encapsulation Mechanism (KEM) built from Niederreiter's dual version of McEliece's PKE (Public-Key Encryption) using binary Goppa codes.
- CRYSTALS-Kyber [113]: it is a cryptographic suite that consists of two primitives: a KEM (Kyber) and a digital signature algorithm (Dilithium). Such primitives security is based on the hardness of solving the LWE problem over module lattices.
- FrodoKEM [114]: it is a KEM whose security relies on the hardness of the LWE problem. It was developed in part by researchers from Microsoft, Google and NXP.

- HQC (Hamming Quasi-Cyclic) [115]: it is a code-based PKE scheme that uses quasi-cyclic codes as well as BCH (Bose–Chaudhuri–Hocquenghem) codes.
- LAC [116]: it is a lattice-based post-quantum PKE that relies on the hardness of the Ring-LWE problem.
- LEDACrypt [117]: it is a cryptographic suite built on QC-LDPC codes that includes both a KEM and a Public Key Cryptosystem (PKC).
- New Hope [118]: it is a lattice-based post-quantum scheme that relies on the Ring-LWE problem.
- NTRU Encrypt [119]: it is a lattice-based cryptosystem based on the hardness on the LWE and Ring-LWE problems.
- NTRU Prime [120]: it is a Ring-LWE based cryptosystem that tweaks NTRU to improve its security. This improvement is achieved by using rings but avoiding the use of certain special structures that have been exploited to perform certain recent attacks.
- NTS-KEM [121]: it is a code-based post-quantum cryptosystem based on McEliece and Niederreiter schemes.
- ROLLO-II [122]: it is a code-based cryptosystem based on the use of rank metric and LRPC codes.
- Round-5 [123]: it is a lattice-based cryptosystem that relies on the General Learning with Rounding (GLWR) problem to unify the well-studied Learning with Rounding (LWR) and Ring Learning with Rounding (RLWR) lattice-problems.
- RQC [124]: it is a code-based PKE that uses both ideal and Gabidulin codes.
- SABER [125]: it is a KEM whose security that relies on the hardness of the Module Learning With Rounding problem (MLWR).
- SIKE [126]: it is an isogeny-based key encapsulation suite based on pseudo-random walks in supersingular isogeny graphs.
- Three Bears [127]: it is a lattice-based cryptosystem based on integer module learning with errors (I-MLWE).

The algorithm list in Table II includes variants of the previously mentioned 17 second-round NIST cryptosystems that provide different security levels and that currently seem more promising for being implemented in IoT devices due to their computational complexity and required key sizes. Nonetheless, it is important to note that most of such systems can be tweaked to reach a trade-off between security level and performance. The following are the main conclusions that can be drawn after analyzing the Table:

- Most candidates are either code-based or lattice-based. There are no multivariate-based candidates and only one isogeny-based cryptosystem (SIKE).
- Most code-based candidates make use of quasi-cyclic codes or are based on McEliece or Niederreiter schemes. Regarding the lattice-based cryptosystems, the majority are based on solving the LWE or the Learning with Rounding (LWR) problem (or one of their variants).
- The cryptosystems that indicate explicitly their quantum security level (which quantifies the computational effort required to break the cryptosystem with a quantum com-puter), estimate that they provide between 64 and 308 bits, so they seem secure for the near and middle term. Regarding the claimed security level against classical computing attacks, the algorithms can be classified into three categories indicated by NIST: Category 1 is for a 128-bit security level (i.e., it is equivalent to break AES-128), Category 3 is for 192 bits and Category 5 is for 256 bits.

- The last columns of Table II can be a good reference for developers when selecting a post-quantum algorithm to be implemented in an IoT system, since key sizes are essential for determining the required memory and computational resources:
  - As it can be observed, the majority of the cryptosystems make use of key sizes larger than the ones required by classical asymmetric and symmetric cryptosystems. However, there is a wide range of algorithms that use between 2,640 (SIKE) and 11,357,632 (NTS-KEM) bits for a public key, and between 128 (Round5 KEM for IoT) and 159,376 (NTS-KEM) bits for a private key. It is worth noting that, although in some cryptosystems like NTS-KEM public keys are really large, the length of the ciphertext they generate is short (e.g., 2,024 bits for the highest 256-bit security level).
  - In the case of the private key it is important to note that some algorithms indicate two sizes: the one inside parentheses is much smaller, since it is the resulting size of using a seed expander (therefore, in practice, the key will occupy the amount of memory indicated outside of the parentheses). In the particular cases of FrodoKEM, the size indicated for the private key is actually the sum of the private key size and the public key size.
  - Since, the smaller the key sizes, the easier the implementation in resource-constrained devices, only a few cryptosystems like Round5 or SIKE seem really promising for the current IoT end-node hardware.
- Finally, the last column of Table II indicates the most relevant references, where further details on the algorithms can be obtained.

## V. Security in IoT communications architectures

It is important to note that post-quantum security goes beyond IoT node security and should be regarded as a whole, securing all the elements involved in an IoT communications architecture. This is related to the fact that, nowadays, due to the computational constraints of IoT end devices, the most complex processing tasks are usually carried out in centralized servers or clouds on the Internet [128]. As an example, Figure 2 depicts a generic three-layer cloud-based IoT architecture. The layer at the bottom includes all the IoT nodes, which can be organized in topologies like stars or mesh networks. Every IoT node is able to reach the cloud by using IoT routing/coordinator nodes and/or intermediate gateways. At the top of Figure 2 is the cloud, which provides access to remote users, to other IoT networks or to third-party services, which can either access the cloud or provide services to it.

TABLE II
COMPARISON OF THE POST-QUANTUM ALGORITHMS THAT PASSED TO THE SECOND ROUND OF THE NIST CALL.

| Cryptosystem | Type | Subtype | Claimed Quantum Security | Claimed Classical Security | Public Key Size (bits) | Private Key Size (bits) | Key References |
|---|---|---|---|---|---|---|---|
| BIKE-1 Level 1 | Code based | QC-MDPC McEliece | - | 128 bits | 20,326 | 2,130 | [111], [129], [130], [131], [132], [133], [134] |
| BIKE-1 Level 3 | Code based | QC-MDPC McEliece | - | 192 bits | 39,706 | 3,090 | [111], [129], [130], [131], [132], [133], [134] |
| BIKE-1 Level 5 | Code based | QC-MDPC McEliece | - | 256 bits | 65,498 | 4,384 | [111], [129], [130], [131], [132], [133], [134] |
| BIKE-2 Level 1 | Code based | QC-MDPC Niederreiter | - | 128 bits | 10,163 | 2,130 | [111], [129], [130], [131], [132], [133], [134] |
| BIKE-2 Level 3 | Code based | QC-MDPC Niederreiter | - | 192 bits | 19,853 | 3,090 | [111], [129], [130], [131], [132], [133], [134] |
| BIKE-2 Level 5 | Code based | QC-MDPC Niederreiter | - | 256 bits | 32,749 | 4,384 | [111], [129], [130], [131], [132], [133], [134] |
| BIKE-3 Level 1 | Code based | QC-MDPC Ouroboros | - | 128 bits | 22,054 | 2,010 | [111], [129], [130], [131], [132], [133], [134] |
| BIKE-3 Level 3 | Code based | QC-MDPC Niederreiter | - | 192 bits | 43,366 | 2,970 | [111], [129], [130], [131], [132], [133], [134] |
| BIKE-3 Level 5 | Code based | QC-MDPC Niederreiter | - | 256 bits | 72,262 | 4,256 | [111], [129], [130], [131], [132], [133], [134] |
| Classic McEliece (mceliece8192128) CPU based | Code based | Niederreiter's dual version using binary Goppa codes | - | 256 bits | 10,862,592 | 112,640 | [112], [135], [136], [137] |
| CRYSTALS Kyber-512 | Lattice based | Based on solving the LWE problem with Module Lattices | 100 bits | 128 bits | 6,400 | 13,056 (256) | [113], [138], [139], [140] |
| CRYSTALS Kyber-512 90s | Lattice based | Based on solving the LWE problem with Module Lattices | 100 bits | 128 bits | 6,400 | 13,056 (256) | [113], [138], [139], [140] |
| CRYSTALS Kyber-768 | Lattice based | Based on solving the LWE problem with Module Lattices | 164 bits | 192 bits | 9,472 | 19,200 (256) | [113], [138], [139], [140] |
| CRYSTALS Kyber-768 90s | Lattice based | Based on solving the LWE problem with Module Lattices | 164 bits | 192 bits | 9,472 | 19,200 (256) | [113], [138], [139], [140] |
| CRYSTALS Kyber-1024 | Lattice based | Based on solving the LWE problem with Module Lattices | 230 bits | 256 bits | 12,544 | 25,344 (256) | [113], [138], [139], [140] |
| CRYSTALS Kyber-1024 90s | Lattice based | Based on solving the LWE problem with Module Lattices | 230 bits | 256 bits | 12,544 | 25,344 (256) | [113], [138], [139], [140] |
| FrodoKEM-640 AES | Lattice based | Based on solving the LWE problem with generic "algebraically unstructured" lattices | - | 128 bits | 76,928 | 159,104 | [90], [114], [141], [142] |
| FrodoKEM-640 SHAKE | Lattice based | Based on solving the LWE problem with generic "algebraically unstructured" lattices | - | 128 bits | 76,928 | 159,104 | [90], [114], [141], [142] |
| FrodoKEM-976 AES | Lattice based | Based on solving the LWE problem with generic "algebraically unstructured" lattices | - | 192 bits | 125,056 | 250,368 | [90], [114], [141], [142] |
| FrodoKEM-976 SHAKE | Lattice based | Based on solving the LWE problem with generic "algebraically unstructured" lattices | - | 192 bits | 125,056 | 250,368 | [90], [114], [141], [142] |
| FrodoKEM-1344 AES | Lattice based | Based on solving the LWE problem with generic "algebraically unstructured" lattices | - | 256 bits | 172,160 | 344,704 | [90], [114], [141], [142] |
| FrodoKEM-1344 SHAKE | Lattice based | Based on solving the LWE problem with generic "algebraically unstructured" lattices | - | 256 bits | 172,160 | 344,704 | [90], [114], [141], [142] |
| HQC Level 1 (hqc-128-1) | Code based | Quasi-Cyclic and BCH codes | 64 bits | 128 bits | 49,360 (25,000) | 2016 (320) | [115], [143], [144] |
| HQC Level 3 (hqc-192-1) | Code based | Quasi-Cyclic and BCH codes | 96 bits | 192 bits | 87,344 (43,992) | 3,232 (320) | [115], [143], [144] |
| HQC Level 5 (hqc-256-1) | Code based | Quasi-Cyclic and BCH codes | 128 bits | 256 bits | 127,184 (63,912) | 4,256 (320) | [115], [143], [144] |
| LAC-128 (CCA) | Lattice based | Based on solving Ring-LWE | - | 128 bits | 4,352 | 8,448 | [116] |
| LAC-192 (CCA) | Lattice based | Based on solving Ring-LWE | - | 192 bits | 8,448 | 16,640 | [116] |
| LAC-256 (CCA) | Lattice based | Based on solving Ring-LWE | - | 256 bits | 8,448 | 16,640 | [116] |
| LEDACrypt KEM Level 1 (for two circulant blocks) | Code based | QC-LDPC Niederreiter | - | 128 bits | 14,976 | 3,616(192) | [117], [145], [146] |
| LEDACrypt KEM Level 3 (for two circulant blocks) | Code based | QC-LDPC Niederreiter | - | 192 bits | 25,728 | 5,152 (256) | [117], [145], [146] |
| LEDACrypt KEM Level 5 (for two circulant blocks) | Code based | QC-LDPC Niederreiter | - | 256 bits | 36,928 | 6,112 (320) | [117], [145], [146] |
| NewHope-512 (CCA) | Lattice based | Based on solving Ring-LWE | 101 bits | 128 bits | 7,424 | 15,104 | [92], [118], [147], [148] |
| NewHope-1024 (CCA) | Lattice based | Based on solving Ring-LWE | 233 bits | 256 bits | 14,592 | 29,440 | [92], [118], [147], [148] |
| NTRUEncrypt-443 | Lattice based | Based on solving LWE/Ring-LWE | 84 bits | 128 bits | 4,888 | 5,608 | [85], [119], [149] |
| NTRUEncrypt-743 | Lattice based | Based on solving LWE/Ring-LWE | 159 bits | 192 bits | 8,184 | 9,384 | [85], [119], [149] |
| NTRUEncrypt-1024 | Lattice based | Based on solving LWE/Ring-LWE | 198 bits | 256 bits | 32,776 | 65,552 | [85], [119], [149] |
| NTRU Prime (sntrup4591761) | Lattice based | Based on solving Ring-LWE | 139-208 bits | 153-368 bits | 9,744 | 12,800 | [120], [150], [151] |
| NTRU Prime (ntrulpr4591761) | Lattice based | Based on solving Ring-LWE | 140-210 bits | 155-364 bits | 8,376 | 9,904 | [120], [150], [151] |
| NTS-KEM Level 1 | Code based | Based on McEliece and Niederreiter | 64 bits | 128 bits | 2,555,904 | 73,984 | [121], [152] |
| NTS-KEM Level 3 | Code based | Based on McEliece and Niederreiter | 96 bits | 192 bits | 7,438,080 | 140,448 | [121], [152] |
| NTS-KEM Level 5 | Code based | Based on McEliece and Niederreiter | 128 bits | 256 bits | 11,357,632 | 159,376 | [121], [152] |
| ROLLO-II 128 | Code based | Based on rank metric codes with LRPC codes | - | 128 bits | 12,368 | 320 | [122], [153] |
| ROLLO-II 192 | Code based | Based on rank metric codes with LRPC codes | - | 192 bits | 16,160 | 320 | [122], [153] |
| ROLLO-II 256 | Code based | Based on rank metric codes with LRPC codes | - | 256 bits | 19,944 | 320 | [122], [153] |
| Round5 KEM IoT | Lattice based | Based on General Learning with Rounding (GLWR) | 88-101 bits | 96-202 bits | 2,736 | 128 | [123], [154] |
| RQC-I | Code based | Based on Rank Quasi-Cyclic codes | - | 128 bits | 6,824 | 320 | [144], [124], [155] |
| RQC-II | Code based | Based on Rank Quasi-Cyclic codes | - | 192 bits | 11,128 | 320 | [144], [124], [155] |
| RQC-III | Code based | Based on Rank Quasi-Cyclic codes | - | 256 bits | 18,272 | 320 | [144], [124], [155] |
| SABER KEM (LightSABER) | Lattice based | Based on Module Learning With Rounding problem (Mod-LWR) | 114-153 bits | 125-169 bits | 5,376 | 12,544 ( 7,936) | [125], [156] |
| SABER KEM | Lattice based | Based on Module Learning With Rounding problem (Mod-LWR) | 185-226 bits | 203-244 bits | 7,936 | 18,432 (10,752) | [125], [156] |
| SABER KEM (FireSABER) | Lattice based | Based on Module Learning With Rounding problem (Mod-LWR) | 257-308 bits | 283-338 bits | 10,496 | 24,320 (14,080) | [125], [156] |
| SIKE (SIKEp434) | Isogeny based | Based on pseudo-random walks in supersingular isogeny graphs | - | 128 bits | 2,640 | 2,992 | [126], [157] |
| Three Bears (BabyBear CCA) | Lattice based | Based on integer module learning with errors (I-MLWE) | 140-180 bits | 154-190 bits | 6,432 | 320 | [127] |
| Three Bears (MamaBear CCA) | Lattice based | Based on integer module learning with errors (I-MLWE) | 213-228 bits | 235-241 bits | 9,552 | 320 | [127] |
| Three Bears (PapaBear CCA) | Lattice based | Based on integer module learning with errors (I-MLWE) | 285-300 bits | 314-317 bits | 12,672 | 320 | [127] |

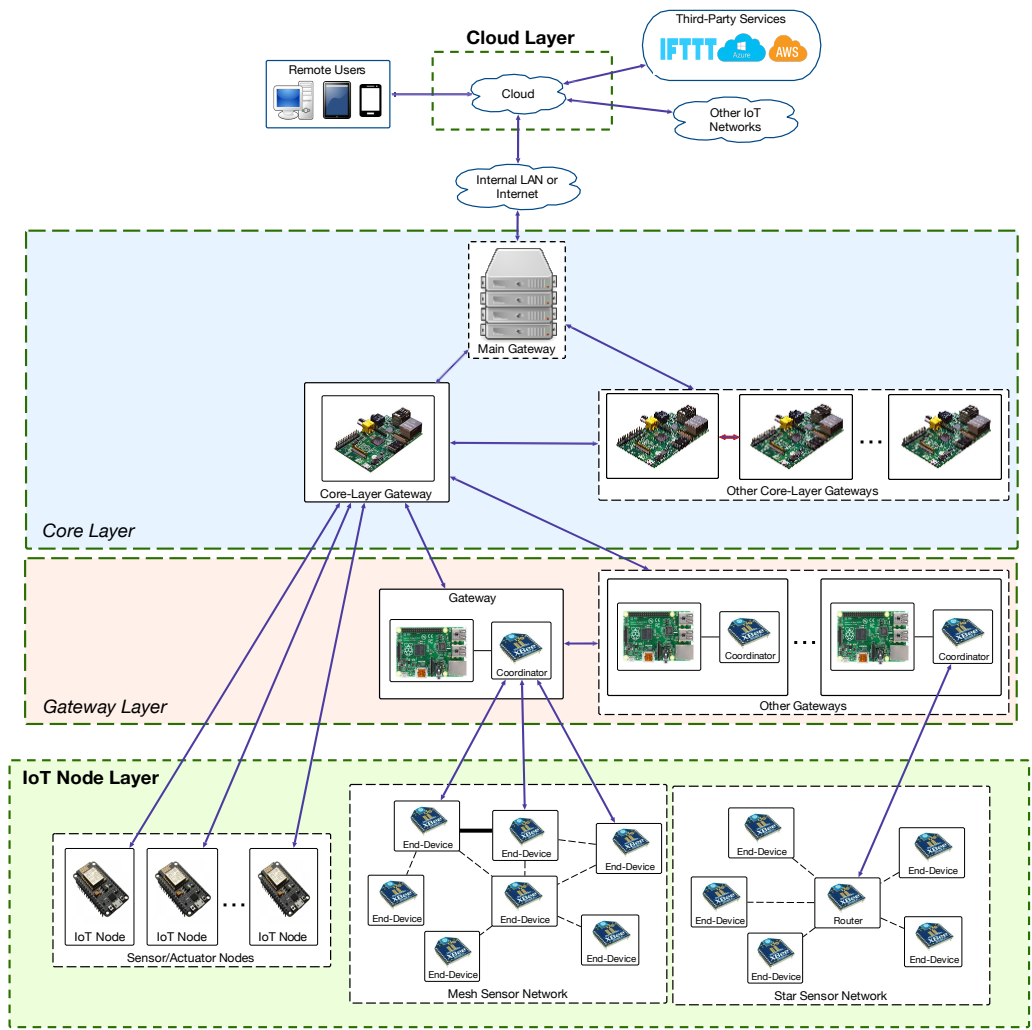

Fig. 2. Traditional cloud-based IoT architecture.

However, cloud-based systems have certain limitations when dealing with large-scale IoT deployments [158], so academia and industry are currently exploring new paradigms like Edge, Fog or Mist computing [159], [160], [161] in order to develop new IoT architectures. An example of Edge architecture is shown in Figure 3, which is composed by four essential layers:

- *ì* IoT Node Layer. It is the layer at the bottom and is composed by IoT nodes and actuators. Such IoT nodes exchange information directly with gateways or with other IoT nodes, usually conforming a mesh network.
- *ì* IoT Node Gateway Layer. Certain IoT nodes, due to their communication range, energy consumption constraints or supported protocols, need to make use of intermediate gateways before reaching the Edge Layer. This is for instance the case of certain heterogeneous sensor networks that make use of multiple communications protocols [163].
- *ì* Edge Layer. In contrast to the Gateway Layer of the traditional IoT architecture represented in Figure 2, the

Edge Layer in Figure 3 is not only aimed at routing data, but it provides advanced edge computing services (e.g., sensor fusion or fog services) through the use of cloudlets and fog computing nodes [164]. Specifically, the Edge Layer is composed by two sublayers: the Fog Sublayer and the Cloudlet Sublayer. The Fog Sublayer essentially consists of fog computing nodes, which are ideal for providing physically-distributed and low-latency IoT applications [159], [165], but, since they are usually constrained in terms of computing power [166], cloudlets provide support for compute-intensive tasks [167], [164].

- *ì* Cloud. It is at the top of the architecture and, like in the architecture depicted in Figure 2, it provides access to remote users, to other remote management software and to third-party services.

There are other architectural paradigms and among them, it is worth mentioning the decentralized approach provided by blockchain-based architectures, which seem to be one of the most promising due to their ability to provide redundancy, attack resilience and scalability [168], [169], [170], [171],

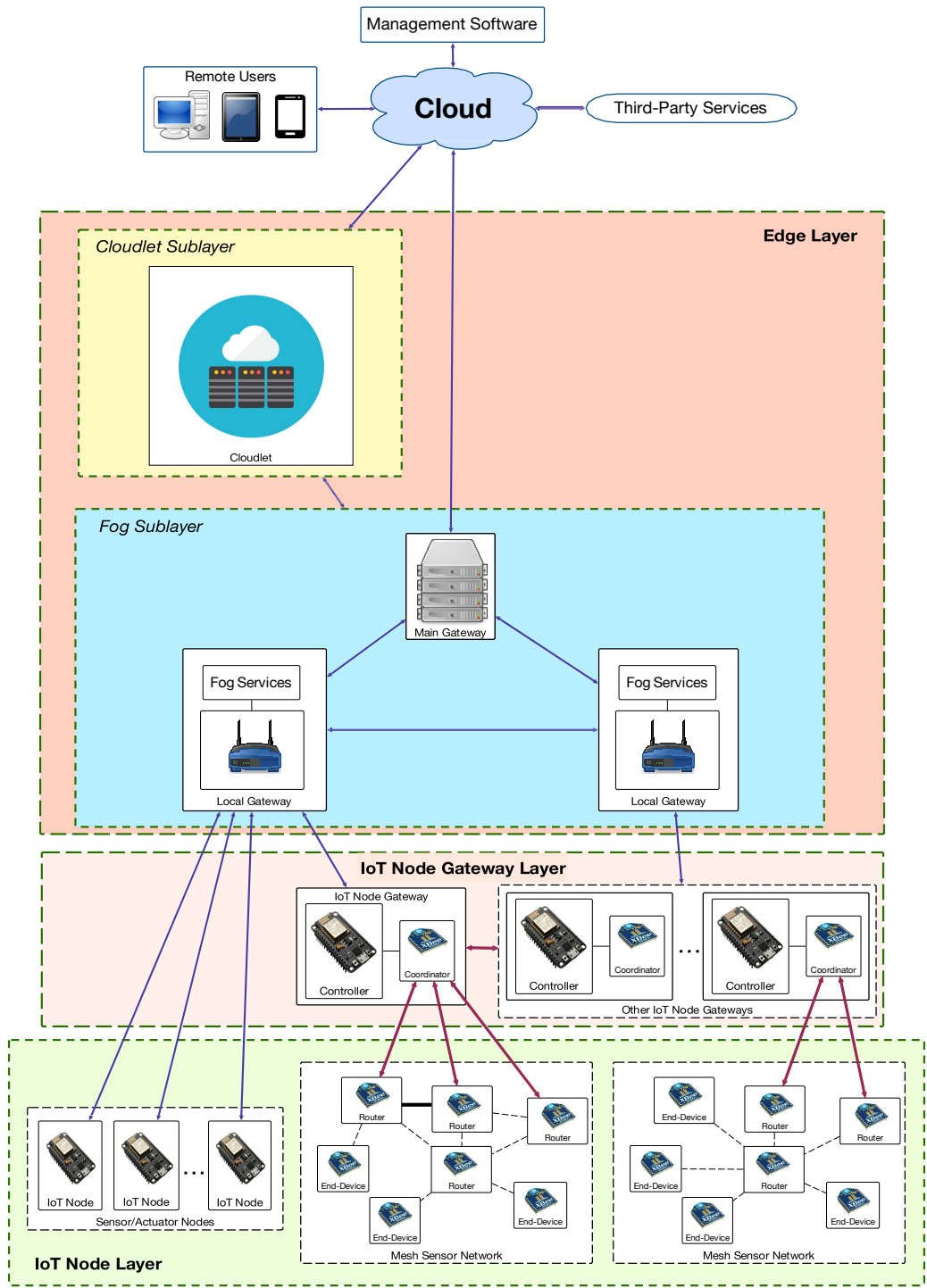

Fig. 3. Traditional cloud-based IoT architecture.

[172], [173].

In any case, when developing or adapting post-quantum cryptosytems for its use in IoT architectures, for the sake of fairness, it is necessary to evaluate the whole architecture in terms of performance and energy efficiency. Such an evaluation essentially involves assessing the performance of the cloud, edge computing devices (e.g., fog gateways, mist nodes, cloudlets) and the communications protocols they use. Specifically, there are five main types of IoT communications that need to be secured:

ⅰ Communications among IoT nodes.

ⅰ Communications between an IoT node and an IoT gate-way.
- Communications among IoT gateways or Edge computing devices.
- Communications between an IoT gateway and the cloud.
- Communications between an IoT node and the cloud.

Therefore, post-quantum security affects the whole IoT architecture, since it is essential for the encryption and authentication of end-point IoT devices, for the network infrastructure encryption or for preserving the security of cloud storage/computing, Big Data, data mining or the exchanged machine learning information [45]. Since the computational power of the hardware involved in the different layers of a communications architecture varies from one layer to another, it is essential to evaluate the requirements of such hardware and choose the appropriate post-quantum protocols and algorithms.

Thus, the next section analyzes the performance of post-quantum cryptosystems when they are executed by resource-constrained devices (that may belong to the IoT Node Layer), medium-power devices (with a power similar to the current fog gateways of the Edge Layer) and powerful devices (like the ones required by cloud servers or Cloudlets).

## VI. POST-QUANTUM IMPLEMENTATION PERFORMANCE

### A. Implementations for resource-constrained IoT devices

Only a few recent works have proposed post-quantum solutions specific to resource-constrained IoT devices (many of them are related to the Horizon 2020 project PQCrypto).

In the case of [174], four different ARM Cortex M0 NTRU-Encrypt implementations for four security levels (112, 128, 192 and 256 bits) are described, compared with other similar developments [175], [176], [177], [178] and analyzed (in terms of performance, memory footprint and energy consumption), confirming that they are suitable for battery-operated devices.

In [175] NTRUEncrypt is optimized and evaluated on different embedded devices that are currently outdated, but whose performance is similar to current low-power IoT nodes: a Palm Computing Platform (MC68EX328 Dragonball, Palm Vx), an Intel 60386 microprocessor (RIM 957) and a 37 MHz ARM7-based device. Other NTRUEncrypt evaluations have also been carried out on ATmega128 and ATmega163 microcontrollers [177]. In addition, in [179] variants of Ring-LWE are implemented and compared for an 8-bit Atmel ATxmega128A1 microcontroller. Atmel microcontrollers have been also used for evaluating Ring-LWE [180], [181], QC-MDPC [182] or McEliece [133], [183]. Furthermore, in [184], ARM Cortex M0 microcontrollers are evaluated when executing different versions of Identity-Based Encryption (IBE). The same microcontroller was used for evaluating New Hope [185] and SABER [125].

The performance of the algorithms previously mentioned in this section is summarized in Table III. For the sake of fairness, the shown performance results are compared for similar hardware that can actually be considered IoT platforms, thus disregarding implementations for outdated hardware. Specifically, Table III shows, for every evaluated cryptosystem, the number of cycles consumed by the key generation, encryption and decryption processes when running on a specific embedded hardware platform. Such platforms are mainly low power microcontrollers (either 32-bit (ARM Cortex-M0) or 8-bit (ATxmega128, ATxmega192, ATxmega256) platforms), whose characteristics are shown in Table IV.

It is worth mentioning that the missing values of Table III are due to the fact that, for certain implementations, key generation is performed offline or its execution time was not reported in the referenced work. In addition, in the case of New Hope, results are indicated just as a reference, being related to the key exchange protocol (KEX) implementation described in [185], which differs from the New Hope KEM presented to the NIST call [118]. Thus, it must be noted that the number of cycles provided for the "Key Generation" of the KEX actually refers to the number of cycles required to execute the key exchange at the server and client sides.

The performance differences due to the selected evaluation hardware platform are significant, so developers should take such a factor into account when comparing the number of cycles among algorithms. For instance, among the cryptosystems in Table III that were evaluated with similar ARM Cortex-M0 platforms and that provide accurate measurements for the three compared performance variables, NTRUEncrypt and two of the lightweight Ring-LWE implementations described in [181] are the fastest ones. However, note that such four implementations differ in their security level, providing NTRUEncrypt the highest (128 bits). Regarding the implementations based on 8-bit AVR microcontrollers, it seems that the fastest implementation is the Ring-LWE variant described in [180].

### B. Implementations for fog/edge computing node-like hardware

Some researchers have previously analyzed the performance of post-quantum algorithms in devices whose performance is similar to the one expected from a fog/edge node, which is usually implemented in Single-Board Computers (SBC) like Raspberry Pi or Orange Pi PC [161], [162], [165], [166], [164].

For instance, NTRUEncrypt performance has been evaluated on a smart card with a 32-bit ARM7TDMI [178] and on ARMv6/ARMv7-A/ARMv8-A MPSoCs (Multi-Processor System-of-Chip) [186]. In addition, in [187] it is presented a multi-mode crypto-coprocessor that supports NTRU and TTS, and that is verified on an ARM9-based board (ARM926EJ-S).

The performance of ARM Cortex M4, A72 and A75 devices has also been measured when executing IBE [184], New Hope [185], Kyber [188], FrodoKEM [141], Round5 [154], SABER [125], SIKE [157] and Three Bears [127].

Table V compares the performance of different relevant post-quantum cryptosystems for the previously mentioned hardware platforms. Like in Table III, for every evaluated cryptosystem and platform, it is indicated the number of cycles consumed by the key generation, encryption and decryption processes. For the sake of completeness, Table VI shows the specifications of the compared hardware platforms.

Like in the previous IoT node performance comparison, differences can be observed due to the evaluation hardware. Among the cryptosystems tested on ARM Cortex-M4, Round5 (KEM IoT version) and the variants of CRYSTALS-Kyber and ThreeBears of the Table that provide the lowest security level are the fastest ones. Regarding the cryptosystems tested on more powerful ARMv8-based devices, the 128-bit FrodoKEM-AES implementation is the lightest one, but it still requires many more cycles than the fastest version for an ARM Cortex-M4. However, if it is compared the actual time required by every device, the most powerful devices are faster than the ones based on ARM Cortex-M4, since their period is significantly shorter. For instance, the period of the tested ARM Cortex-M4 microcontroller is 6 ns, while the period of the tested ARM Cortex-A72 is 0.5 ns. These periods mean, for instance, that, while Round5 (KEM IoT version) requires 2.77 ms for encryption, it only takes 2.04 ms for the same operation using 128-bit FrodoKEM-AES on an ARM Cortex-A72.

TABLE III
PERFORMANCE COMPARISON OF POST-QUANTUM ALGORITHMS ON POTENTIAL IOT NODE PLATFORMS.

| References | Cryptosystem | Platform | Pre-Quantum Security Level | Evaluated version | Key Generation (#Cycles) | Encryption (#Cycles) | Decryption (#Cycles) |
|---|---|---|---|---|---|---|---|
| [174] | NTRUEncrypt | ARM Cortex-M0 | 128 bits | N,p,q = (443,3,2048) | 26,114,818 | 588,044 | 950,371 |
| [174] | NTRUEncrypt | ARM Cortex-M0 | 192 bits | N,p,q = (587,3,2048) | 45,659,315 | 1,040,538 | 1,634,821 |
| [174] | NTRUEncrypt | ARM Cortex-M0 | 256 bits | N,p,q = (743,3,2048) | 71,186,959 | 1,411,557 | 2,377,054 |
| [178] | NTRUEncrypt | ATmega64 | 128 bits | N,p,q = (439,3,2048) | 76,800,000 | 2,008,000 | 1,392,000 |
| [179], [189] | Ring-LWE | ATxmega128A1 | 128 bits | RLWE with N,q = (256, 7681) | - | 874,347 | 215,863 |
| [179], [189] | Ring-LWE | ATxmega128A1 | 256 bits | RLWE with N,q = (512, 12289) | - | 2,196,945 | 600,351 |
| [180] | Ring-LWE | ATxmega128A1 | 128 bits | RLWE with N,q = (256, 7681) | 589,900 | 671,628 | 275,646 |
| [180] | Ring-LWE | ATxmega128A1 | 256 bits | RLWE with N,q = (512, 12289) | 2,165,239 | 2,617,459 | 686,367 |
| [182] | QC-MDPC McEliece | ATxmega256A3 | 80 bits | n0 = 2, n = 9600, r = 4800, w = 90, t = 84 | - | 26,767,463 | 86,874,388 |
| [133] | McEliece | ATxmega256A3 | 80 bits | m=3, n=768, k=432, s=$2^4$ t=7 | - | 4,171,734 | 14,497,587 |
| [183] | McEliece | ATxmega192A1 | 80 bits | n = 2048, k = 1751, t = 27 | - | 14,406,080 | 19,751,094 |
| [190] | McEliece Based on Quasi-dyadic Goppa Codes | ATxmega256A3 | 80 bits | n = 2304, k = 1280, t = 64 | - | 6,358,400 | 33,536,000 |
| [184] | Identity-Based Encryption (IBE) | ARM Cortex-M0 | 80 bits | n=512, q = 16,813,057 | - | 3,297,380 | 1,155,000 |
| [184] | Identity-Based Encryption (IBE) | ARM Cortex-M0 | 192 bits | n=1024, q = 134,348,801 | - | 6,202,910 | 2,171,000 |
| [181] | Ring-LWE | ATxmega128A1 | 94 bits | RLWE with (N,q) = (256, 128) | - | 1,573,000 | 740,000 |
| [181] | Ring-LWE | ATxmega128A1 | 84 bits | RLWE with (N,q) = (256,256) | - | 1,507,000 | 700,000 |
| [181] | Ring-LWE | ATxmega128A1 | 190 bits | RLWE with (N,q) = (512,256) | - | 5,899,000 | 2,791,000 |
| [181] | Ring-LWE | ARM Cortex-M0 | 94 bits | RLWE with (N,q) = (256, 128) | - | 999,000 | 437,000 |
| [181] | Ring-LWE | ARM Cortex-M0 | 84 bits | RLWE with (N,q) = (256,256) | - | 944,000 | 403,000 |
| [181] | Ring-LWE | ARM Cortex-M0 | 190 bits | RLWE with (N,q) = (512,256) | - | 3,483,000 | 1,701,000 |
| [125] | SABER | ARM Cortex-M0 | 192 bits | SABER version optimized for memory | 4,786,000 | 6,328,000 | 7,509,000 |
| [185] | New Hope | ARM Cortex-M0 | 256 bits | 1024-NTT (Number-Theoretic Transform) KEX version | 1,467,101 (server) / 1,760,837 (client) | - | - |

TABLE IV
MAIN CHARACTERISTICS OF THE LOW-POWER HARDWARE TESTED IN THE LITERATURE.

| References | Platform | Part Number | Instruction Set | Clock Frequency | Flash Memory | RAM Memory | EEPROM |
|---|---|---|---|---|---|---|---|
| [174], [184], [181] | ARM Cortex-M0 | Infineon XMC1100 | 32 bits | 32 MHz | 64 KB | 16 KB | 8 KB |
| [178] | AVR ATmega | ATmega64 | 8 bits | 16 MHz | 64 KB | 4 KB | 2 KB |
| [179], [189], [181] | AVR ATxmega | ATxmega128A1 | 8 bits | 32 MHz | 128 KB | 8 KB | 2 KB |
| [182], [133], [190] | AVR ATxmega | ATxmega256A3 | 8 bits | 32 MHz | 256 KB | 16 KB | 4 KB |
| [183] | AVR ATxmega | ATxmega192A1 | 8 bits | 32 MHz | 192 KB | 16 KB | 2 KB |

In any case, the recommendation for future IoT developers is to make use of Tables III and V to choose a post-quantum algorithm according to the required security level, the characteristics of the target IoT hardware platform and the needed performance.

### C. FPGA implementations

FPGAs have been used in the past for carrying out complex processes [191], [192], so they may not be considered low-power IoT devices, but it is interesting to consider different implementations carried out in the past by some researchers. Specifically, Table VII compares the performance of the most relevant post-quantum cryptosystem implementations on FP-GAs. For the sake of fairness, all the included implementations are for Xilinx FPGAs. No FPGA implementations were found for other traditional FPGA manufacturers like Lattice or Actel (now Microsemi), but at least one implementation was found for an Altera (now Intel) FPGA (an Stratix V 5SGXEA7N) [135].

Among the evaluated FPGAs, there are outdated models (e.g., the Xilinx Virtex-E series, released in the late 90s) to the latest Virtex-7 (from 2010), so there are significant difference in their power consumption, performance, maximum clock frequency and internal logic resource structure (e.g., in the number and location of flip-flops (FFs), Look-Up Tables (LUTs), slices and blocks of RAM (BRAMs)).

Despite the mentioned hardware differences, Table VII shows that certain post-quantum algorithms can be executed really fast in an FPGA. For instance, in [175] and [193] it is evaluated NTRUEncrypt on old FPGAs (a Virtex 1000EFG860 and a Virtex XCV1600E), but their encryption/decryption times are really low (less than 6 $\mu$s). Low execution times

(roughly 8.1 $\mu$s per operation) are also achieved in [194] with a Virtex-6 for a Ring-LWE implementation.

On the opposite side, there are implementations that require more then 1 ms for encryption/decryption. For instance, in [183] an implementation of McEliece is detailed that requires 1.079 ms for encryption and 10.726 ms for decryption. These times seem to be related to the used FPGA (a Xilinx Spartan-3AN, which is a currently outdated low-cost FPGGA), since there are similar implementations of McEliece [182] that use newer models (Virtex-6 XC6VLX240T) that can execute encryption in 13.66 ms and decryption in 85.79 ms.

Another quite slow implementation (in comparison to other post-quantum cryptosystems in Table VII) is presented in [195], which evaluates SIDH in a relatively new and fast FPGA (Virtex-7), requiring 33.7 ms for encryption and 51.4 ms for decryption. In this case the performance is related to the evaluated post-quantum cryptosystem, since there are other developments in the literature that achieve clearly lower execution times with less advanced FPGAs. For example, there are different faster implementations of Ring-LWE and IBE for Virtex-6 and Spartan-6 FPGAs [184], [196], [197], [198], or for a Virtex-7, but for other cryptosystems like SIKE [157].

### D. Implementations for the cloud and cloudlets

Table VIII compares the cryptosystems previously listed in Table II, but in terms of their performance when executed on hardware platforms that have the power similar to cloud servers or cloudlets.

For the sake of fairness, Table VIII shows the performance obtained when running the algorithms in an Intel's x64 architecture, although the version of the microprocessor varies from one to another. The main characteristics of such microprocessors are summarized in Table IV, which shows that there are significant differences among them in terms of clock frequency, main target platform (the microprocessors are designed for Desktop PC, laptops or servers), energy consumption (expressed as Thermal Design Power (TDP)) and performance (according to the Passmark CPU benchmarks [199]). In addition, it is worth pointing out that all the performance results shown in Table VIII were obtained with Turbo Boost and Hyper-Threading disabled, and that it is possible to optimize the algorithms (in fact, many of them have already been optimized for different instruction sets, like the Advanced Vector Extensions 2 (AVX2), a 256-bit instruction set provided by Intel).

The performance results show the number of cycles required by each microprocessor for three essential tasks performed by the cryptosystem: the generation of a key, encapsulation/encryption and decapsulation/decryption. In the case of

TABLE V

PERFORMANCE COMPARISON OF POST-QUANTUM ALGORITHMS ON POTENTIAL IoT FOG/EDGE COMPUTING PLATFORMS.

| References | Cryptosystem | Platform | Pre-Quantum Security Level | Evaluated version | Key Generation (#Cycles) | Encryption (#Cycles) | Decrypton (#Cycles) |
|---|---|---|---|---|---|---|---|
| [178] | NTRUEncrypt | ARM7TDMI | 128 bits | N,p,q = (439,3,2048) | 26,469,600 | 693,720 | 998,760 |
| [184] | Identity-Based Encryption (IBE) | ARM Cortex-M4 | 80 bits | n=512, q = 16,813,057 | - | 972,744 | 318,539 |
| [184] | Identity-Based Encryption (IBE) | ARM Cortex-M4 | 192 bits | n=1024, q = 134,348,801 | - | 1,719,444 | 557,015 |
| [188] | CRYSTALS-Kyber | ARM Cortex-M4 | 100 bits | - | 499,000 | 634,000 | 730,000 |
| [188] | CRYSTALS-Kyber | ARM Cortex-M4 | 164 bits | - | 947,000 | 1,113,000 | 1,059,000 |
| [188] | CRYSTALS-Kyber | ARM Cortex-M4 | 230 bits | - | 1,525,000 | 1,732,000 | 1,653,000 |
| [141] | FrodoKEM-AES | ARM Cortex-A72 | 128 bits | D = 15, q = 32768, n = 640 | 3,470,000 | 4,057,000 | 3,969,000 |
| [141] | FrodoKEM-AES | ARM Cortex-A72 | 192 bits | D = 16, q = 65536, n = 976 | 7,219,000 | 8,530,000 | 8,014,000 |
| [141] | FrodoKEM-AES | ARM Cortex-A72 | 256 bits | D = 16, q = 65536, n = 1344 | 12,789,000 | 14,854,000 | 14,635,000 |
| [141] | FrodoKEM-SHAKE | ARM Cortex-A72 | 128 bits | D = 15, q = 32768, n = 640 | 11,278,000 | 12,411,000 | 12,311,000 |
| [141] | FrodoKEM-SHAKE | ARM Cortex-A72 | 192 bits | D = 16, q = 65536, n = 976 | 24,844,000 | 27,033,000 | 26,936,000 |
| [141] | FrodoKEM-SHAKE | ARM Cortex-A72 | 256 bits | D = 16, q = 65536, n = 1,344 | 44,573,000 | 48,554,000 | 48,449,000 |
| [147] | New Hope-1024 (CCA) | ARM Cortex-M4 | 256 bits | 1024 CCA-KEM version | 1,142,743 | 1,785,343 | 1,797,757 |
| [154] | Round5 | ARM Cortex-M4 | 96-202 bits | KEM IoT version | 341,000 | 465,000 | 191,000 |
| [125] | SABER | ARM Cortex-M4 | 192 bits | SABER version optimized for speed | 1,147,000 | 1,444,000 | 1,543,000 |
| [157] | SIKE | ARM Cortex-A75 | 128 bits | ARMv8 assembly version | 26,928,000 | 44,388,000 | 47,377,000 |
| [127] | ThreeBears | ARM Cortex-M4 | 154-190 bits | BabyBear CCA high-speed version | 644,000 | 824,000 | 1,299,000 |
| [127] | ThreeBears | ARM Cortex-M4 | 235-241 bits | MamaBear CCA high-speed version | 1,257,000 | 1,494,000 | 2,174,000 |
| [127] | ThreeBears | ARM Cortex-M4 | 314-317 bits | PapaBear CCA high-speed version | 2,082,000 | 2,378,000 | 3,272,000 |

TABLE VI
COMPARISON OF THE HARDWARE TESTED FOR EMBEDDED PLATFORMS.

| References | Platform | Part Number | Instruction Set | Clock Frequency | Flash Memory | RAM Memory |
|---|---|---|---|---|---|---|
| [178] | ARMv7 | ARM7TDMI | 32 bits | 4.92 MHz | Not specified (at least 59.3 KB) | Not specified (at least 6 KB) |
| [127], [154], [184], [188] | ARM Cortex-M4 | TM32F407 Discovery kit | 32 bits | 168 MHz | 1 MB | 192 KB |
| [141] | ARMv8-A | ARM Cortex-A72 | 64 bits | 1.992 GHz | Not specified | Not specified |
| [157] | ARMv8 | ARM Cortex-A75 + Cortex-A55 (Google Pixel 3 smartphone) | 64 bits | 4 Cores @  2.45 GHz + 4 Cores @   1.6 GHz | Not specified (64 or 128 GB) | 4 GB |

LEDACrypt, the number of cycles has been obtained by converting the times provided by its NIST second-round submission documentation [145], which indicates that such times were estimated for optimized AVX2-enabled versions of the cryptosystem. In the case of CRYSTALS-Kyber it is indicated inside the parentheses the approximate values when including key generation in decapsulation to avoid having to store expanded private keys.

It can be observed in the Table that there are a few cryptosystems that stand out regarding the low number of cycles they require despite using similar evaluation hardware respect to other algorithms: NTRU Prime, Round5, Three Bears and SABER (LightSABER). However, note that Three Bears, Round 5 and LightSABER were evaluated with low-power microprocessors in laptops, while NTRU Prime was evaluated with a powerful Intel Xeon processor used in servers.

Despite the performance results shown in Table VIII, it must be noted that there are optimized versions of certain algorithms (for instance, SIKE) that are much faster than the 'standard' version, so it is recommended for the interested IoT developer to take a look periodically at the provided reference to keep up with the latest algorithm developments.

## VII. MAIN CHALLENGES AND FUTURE TRENDS IN POST-QUANTUM IoT DEVELOPMENTS

### A. Main challenges

Quantum cryptanalytics and post-quantum cryptography are relatively new fields that are currently being studied and developed by industry, security agencies and academia, hence their foundations are still being established. Such a situation implies that the development of the mentioned fields and their application to IoT involve the significant challenges detailed in the next subsections and illustrated in Figure 4.

*1) Quantum computing fast evolution:* Since quantum computing is currently evolving continuously, it is not guaranteed that the developed IoT post-quantum cryptosystems will resist new algorithms and novel attacks [46]. To mitigate such a threat, it is necessary for IoT developers to pay close attention to the quantum computing scene and its advances.

*2) Large key sizes:* Although most current public-key cryptosystems make use of similar relatively small key sizes (usually between 128 and 4,096 bits), they are often much larger in post-quantum algorithms (this can be observed in Table II), what may be a problem for current resource-constrained devices. In such a case, it will be necessary to look for a trade-off among key size, security level and performance, and, at the same time, devise ways to adapt the cryptosystems and protocols to the key-size requirements. For instance, it is needed to design and to implement energy-efficient post-quantum lattice-based cryptosystems on IoT devices that manage efficiently the storage and operation with large keys.

*3) Slow key generation:* Some post-quantum cryptosystems limit the number of messages signed with the same key to avoid attacks, so they require to generate new keys for each group of signed messages. Therefore, additional computational resources are necessary to manage key generation, what may not be handled efficiently by traditional IoT devices. In this case, it may be necessary to conceive approaches to tweak post-quantum key generation mechanisms to minimize energy consumption.

*4) Excessive time, energy or computational resource consumption:* It is also a concern that new public-key post-quantum algorithms may also consume a relevant amount of time, energy and computational resources when performing encryption, decryption, signing and signature verification tasks. To avoid such problems in practical scenarios, accurate measurements will have to be performed and then the inefficient implementations and cryptosystems will have to be discarded.

*5) Ongoing standardization:* IoT developers may focus on post-quantum cryptosystems that may not be the ones finally chosen by industry/academia to be standardized (as it was mentioned in Section III-B, such a standardization process is an ongoing effort). This may happen because during most security standardizations the focus is placed on security and performance, while energy consumption is often neglected. To limit the impact of this risk and thus be aligned with industry/academia, the output generated from multiples entities that are carrying out standardization initiatives (e.g., NIST, ETSI, IEEE, ISO, ANSI or IETF/IRTF) should be monitored closely.

*6) Lack of standard security level benchmarks:* Since the bits-of-security paradigm does not take into account the security of algorithms against quantum cryptanalysis, post-quantum cryptographers will need to reach a consensus on how to measure the security against quantum attacks and on which key lengths provide an acceptable security level. In fact, although for symmetric cryptosystems it is suggested to double key length to compensate the quadratic speed-up provided by

TABLE VII
PERFORMANCE COMPARISON OF POST-QUANTUM ALGORITHMS ON FPGAs.

| References | FPGA Model | Cryptosystem | Operation | Clock Frequency | Required Time ($\mu s$) | #Cycles | FFs | LUTs | Slices | BRAM |
|---|---|---|---|---|---|---|---|---|---|---|
| [182] | Xilinx Virtex-6 XC6VLX240T | QC-MDPC McEliece | Encryption | 351.3 MHz | 13.66 | 4,800 | 14,426 | 8,856 | 2,920 | 0 |
| [182] | Xilinx Virtex-6 XC6VLX240T | QC-MDPC McEliece | Decryption | 190.6 MHz | 85.79 | 16,352 | 46,515 | 46,249 | 17,120 | 0 |
| [182] | Xilinx Virtex-6 XC6VLX240T | QC-MDPC McEliece | Iterative Decryption | 222.5 MHz | 125.38 | 27,897 | 32,974 | 36,554 | 10,271 | 0 |
| [175] | Xilinx 1000EFG860 | NTRUEncrypt | Encryption | 50.063 MHz | 5.17 | 259 | - | †60,000 | 6,373 | 0 |
| [183] | Xilinx Spartan-3AN XC3S1400AN-5 | McEliece | Encryption | 150 MHz | 1079.19 | 161,878 | 804 | 1,044 | 668 | 3 |
| [183] | Xilinx Spartan-3AN XC3S1400AN-5 | McEliece | Decryption | 85 MHz | 10726.30 | 911,736 | 8,977 | 22,034 | 11,218 | 20 |
| [184] | Xilinx Spartan-6 S6LX25 | IBE (512/16813057) | Encryption | 174 MHz | 80.21 | 13,958 | 6,067 | 7,023 | - | 16 |
| [184] | Xilinx Spartan-6 S6LX25 | IBE (512/16813057) | Decryption | 174 MHz | 54.77 | 9,530 | 6,067 | 7,023 | - | 16 |
| [184] | Xilinx Spartan-6 S6LX25 | IBE (1024/134348801) | Encryption | 174 MHz | 164.29 | 28,586 | 8,686 | 8,882 | - | 27 |
| [184] | Xilinx Spartan-6 S6LX25 | IBE (1024/134348801) | Decryption | 174 MHz | 112.27 | 19,535 | 8,686 | 8,882 | - | 27 |
| [196] | Xilinx Spartan-6 S6LX16 | RLWE | Encryption | 160 MHz | 42.88 | 6,861 | 3,513 | 4,121 | - | 14 |
| [196] | Xilinx Spartan-6 S6LX16 | RLWE | Decryption | 160 MHz | 27.52 | 4,404 | 3,513 | 4,121 | - | 14 |
| [197] | Xilinx Virtex-6 V6LX75T | RWE | Encryption | 278 MHz | 47.84 | 13,300 | 953 | 1,536 | - | 3 |
| [197] | Xilinx Virtex-6 V6LX75T | RWE | Decryption | 278 MHz | 20.86 | 5,800 | 953 | 1,536 | - | 3 |
| [198] | Xilinx Spartan-6 S6LX9 | RLWE | Encryption | 144 MHz | 945.92 | 136,212 | 238 | 282 | - | 1 |
| [198] | Xilinx Spartan-6 S6LX9 | RLWE | Decryption | 189 MHz | 350.99 | 66,338 | 87 | 94 | - | 1 |
| [157] | Xilinx Virtex-7 | SIKE (SIKEp751) | Encryption | 198 MHz | 16.27 | - | 51,914 | 44,822 | 16,756 | 57 |
| [157] | Xilinx Virtex-7 | SIKE (SIKEp751) | Decryption | 198 MHz | 17.08 | - | 51,914 | 44,822 | 16,756 | 57 |
| [195] | Xilinx Virtex-7 | SIDH (Single Core version) | Key Exchange | 177.1 MHz | 33,700 | - | 30,031 | 24,499 | 10,298 | 27 |
| [195] | Xilinx Virtex-7 | SIDH (10-Core version) | Key Exchange | 167.4 MHz | 51,400 | - | 152,134 | 129,784 | 55,838 | 270 |
| [200] | Xilinx Virtex-5 XC5VLX110T | McEliece | Encryption | 163 MHz | 500 | - | - | - | 14,537 | 75 |
| [200] | Xilinx Virtex-5 XC5VLX110T | McEliece | Decryption | 163 MHz | 1,290 | - | - | - | 14,537 | 75 |
| [201] | Xilinx Virtex-6 XC6VLX240T | Niederreiter | Encryption | 300 MHz | 0.66 | 200 | 875 | 926 | 315 | 17 |
| [201] | Xilinx Virtex-6 XC6VLX240T | Niederreiter | Decryption | 250 MHz | 58.78 | 14,500 | 12,861 | 9,409 | 3,887 | 9 |
| [194] | Xilinx Virtex-6 XC6VLX240T | RLWE | Encryption | - | 8.10 | - | 143,396 | 298,016 | - | 0 |
| [194] | Xilinx Virtex-6 XC6VLX240T | RLWE | Decryption | - | 8.15 | - | 65,174 | 124,158 | - | 0 |
| [193] | Xilinx Virtex XCV1600E | NTRUEncrypt | Encryption | 62.3 MHz | 1.54 | 96 | 5,160 | 27,292 | 14,352 | 0 |
| [193] | Xilinx Virtex XCV1600E | NTRUEncrypt | Decryption | 62.3 MHz | 1.41 | 88 | 5,160 | 27,292 | 14,352 | 0 |

TABLE VIII
PERFORMANCE COMPARISON OF POST-QUANTUM ALGORITHMS FOR CLOUDS AND CLOUDLETS.

| References | Cryptosystem | Claimed Classical Security | Performance Evaluation Hardware | Key Generation (#Cycles) | Encapsulation (#Cycles) | Decapsulation (#Cycles) |
|---|---|---|---|---|---|---|
| [111] | BIKE-1 Level 1 | 128 bits | Intel Core i5-6260U @ 1.80 GHz, 32 GB of RAM | 730,025 | 689,193 | 2,901,203 |
| [111] | BIKE-1 Level 3 | 192 bits | Intel Core i5-6260U @ 1.80 GHz, 32 GB of RAM | 1,709,921 | 1,850,425 | 7,666,855 |
| [111] | BIKE-1 Level 5 | 256 bits | Intel Core i5-6260U @ 1.80 GHz, 32 GB of RAM | 2,986,647 | 3,023,816 | 17,483,906 |
| [111] | BIKE-2 Level 1 | 128 bits | Intel Core i5-6260U @ 1.80 GHz, 32 GB of RAM | 6,383,408 | 281,755 | 2,674,115 |
| [111] | BIKE-2 Level 3 | 192 bits | Intel Core i5-6260U @ 1.80 GHz, 32 GB of RAM | 22,205,901 | 710,970 | 7,114,241 |
| [111] | BIKE-2 Level 5 | 256 bits | Intel Core i5-6260U @ 1.80 GHz, 32 GB of RAM | 58,806,046 | 1,201,161 | 16,385,956 |
| [111] | BIKE-3 Level 1 | 128 bits | Intel Core i5-6260U @ 1.80 GHz, 32 GB of RAM | 433,258 | 575,237 | 3,437,956 |
| [111] | BIKE-3 Level 3 | 192 bits | Intel Core i5-6260U @ 1.80 GHz, 32 GB of RAM | 1,100,372 | 1,460,866 | 7,732,167 |
| [111] | BIKE-3 Level 5 | 256 bits | Intel Core i5-6260U @ 1.80 GHz, 32 GB of RAM | 2,300,332 | 3,257,675 | 18,047,493 |
| [112] | Classic McEliece (mceliece8192128) | 256 bits | Intel Xeon E3-1220 v3 @ 3.10 GHz | ↔4,675,000,000 | ←296,000 | ←458,000 |
| [113] | CRYSTALS Kyber-512 | 128 bits | Intel Core i7-4770K @ 3.5 GHz | 118,044 | 161,440 | 190,206 (↑ 279, 150) |
| [113] | CRYSTALS Kyber-512 90s | 128 bits | Intel Core i7-4770K @ 3.5 GHz | 232,368 | 285,336 | 313,452 (↑ 436, 088) |
| [113] | CRYSTALS Kyber-768 | 192 bits | Intel Core i7-4770K @ 3.5 GHz | 217,728 | 272,254 | 315,976 (↑ 469, 008) |
| [113] | CRYSTALS Kyber-768 90s | 192 bits | Intel Core i7-4770K @ 3.5 GHz | 451,018 | 514,088 | 556,972 (↑ 758, 934) |
| [113] | CRYSTALS Kyber-1024 | 256 bits | Intel Core i7-4770K @ 3.5 GHz | 331,418 | 396,928 | 451,096 (↑ 667, 596) |
| [113] | CRYSTALS Kyber-1024 90s | 256 bits | Intel Core i7-4770K @ 3.5 GHz | 735,382 | 810,398 | 860,272 (↑ 1, 148, 394) |
| [114] | FrodoKEM-640 AES | 128 bits | Intel Core i7-6700 @ 3.4 GHz | 1,384,000 | 1,858,000 | 1,749,000 |
| [114] | FrodoKEM-640 SHAKE | 128 bits | Intel Core i7-6700 @ 3.4 GHz | 7,626,000 | 8,362,000 | 8,248,000 |
| [114] | FrodoKEM-976 AES | 192 bits | Intel Core i7-6700 @ 3.4 GHz | 2,820,000 | 3,559,000 | 3,400,000 |
| [114] | FrodoKEM-976 SHAKE | 192 bits | Intel Core i7-6700 @ 3.4 GHz | 16,841,000 | 18,077,000 | 17,925,000 |
| [114] | FrodoKEM-1344 AES | 256 bits | Intel Core i7-6700 @ 3.4 GHz | 4,756,000 | 5,981,000 | 5,748,000 |
| [114] | FrodoKEM-1344 SHAKE | 256 bits | Intel Core i7-6700 @ 3.4 GHz | 30,301,000 | 32,611,000 | 32,387,000 |
| [115] | HQC Level 1 (hqc-128-1) | 128 bits | Intel Core i7- 7820X CPU @ 3.6 GHz, 16 GB of RAM | 110,000 | 190,000 | 310,000 |
| [115] | HQC Level 3 (hqc-192-1) | 192 bits | Intel Core i7- 7820X CPU @ 3.6 GHz , 16 GB of RAM | 190,000 | 330,000 | 510,000 |
| [115] | HQC Level 5 (hqc-256-1) | 256 bits | Intel Core i7- 7820X CPU @ 3.6 GHz , 16 GB of RAM | 270,000 | 470,000 | 690,000 |
| [116] | LAC-128 (CCA) | 128 bits | Intel Core-i7-4770S @ 3.10 GHz, 7.6 GB of RAM | 90,411 | 160,314 | 216,957 |
| [116] | LAC-192 (CCA) | 192 bits | Intel Core-i7-4770S @ 3.10 GHz, 7.6 GB of RAM | 281,324 | 421,439 | 647,030 |
| [116] | LAC-256 (CCA) | 256 bits | Intel Core-i7-4770S @ 3.10 GHz, 7.6 GB of RAM | 267,831 | 526,915 | 874,742 |
| [117] | LEDACrypt KEM Level 1 (for two circulant blocks) | 128 bits | Intel i5-6600 @ 3.6 GHz | 1,062,000,000 | 475,200 | 1,501,200 |
| [117] | LEDACrypt KEM Level 3 (for two circulant blocks) | 192 bits | Intel i5-6600 @ 3.6 GHz | 3,261,600,000 | 932,400 | 3,279,600 |
| [117] | LEDACrypt KEM Level 5 (for two circulant blocks) | 256 bits | Intel i5-6600 @ 3.6 GHz | 9,075,600,000 | 2,466,000 | 5,079,600 |
| [118] | NewHope-512 (CCA) | 128 bits | Intel Core i7-4770K @ 3.5 GHz | 117,128 | 180,648 | 206,244 |
| [118] | NewHope-1024 (CCA) | 256 bits | Intel Core i7-4770K @ 3.5 GHz | 244,944 | 377,092 | 437,056 |
| [119] | NTRUEncrypt (ntruhrss701) | 128/192 bits | Intel Core i7-4770K @ 3.5 GHz | 23,302,424 | 1,256,210 | 3,642,966 |
| [119] | NTRUEncrypt (ntruhps2048677) | 128/192 bits | Intel Core i7-4770K @ 3.5 GHz | 21,833,048 | 1,313,454 | 3,399,726 |
| [119] | NTRUEncrypt (ntruhps4096821) | 192/256 bits | Intel Core i7-4770K @ 3.5 GHz | 31,835,958 | 1,856,936 | 4,920,436 |
| [120] | NTRU Prime (sntrup4591761) | 153-368 bits | Intel Xeon E3-1275 v3 @ 3.5 GHz | 940,852 | 44,788 | 93,676 |
| [120] | NTRU Prime (ntrulpr4591761) | 155-364 bits | Intel Xeon E3-1275 v3 @ 3.5 GHz | 44,948 | 81,144 | 113,708 |
| [121] | NTS-KEM Level 1 | 128 bits | 16-core server with Intel Xeon E5-2667 v2 @ 3.3 GHz, 256 GB of RAM | 39,388,653 | 124,528 | 650,116 |
| [121] | NTS-KEM Level 3 | 192 bits | 16-core server with Intel Xeon E5-2667 v2 @ 3.3 GHz, 256 GB of RAM | 125,672,723 | 396,513 | 1,181,373 |
| [121] | NTS-KEM Level 5 | 256 bits | 16-core server with Intel Xeon E5-2667 v2 @ 3.3 GHz, 256 GB of RAM | 229,357,286 | 532,168 | 2,500,475 |
| [122] | ROLLO-II 128 | 128 bits | Intel Core i7-7820X @ 3.6 GHz, 16 GB of RAM | 9,620,000 | 1,520,000 | 4,960,000 |
| [122] | ROLLO-II 192 | 192 bits | Intel Core i7-7820X @ 3.6 GHz, 16 GB of RAM | 11,040,000 | 2,000,000 | 6,520,000 |
| [122] | ROLLO-II 256 | 256 bits | Intel Core i7-7820X @ 3.6 GHz, 16 GB of RAM | 11,410,000 | 2,390,000 | 7,940,000 |
| [123] | Round5 KEM IoT | 96-202 bits | MacBook Pro 15.1 with Intel Core i7 @ 2.6 GHz | 56,300 | 97,900 | 59,500 |
| [124] | RQC-I | 128 bits | Intel Core i7-7820X @ 3.6 GHz, 16 GB of RAM | 700,000 | 1,300,000 | 6,660,000 |
| [124] | RQC-II | 192 bits | Intel Core i7-7820X @ 3.6 GHz, 16 GB of RAM | 1,120,000 | 2,180,000 | 14,680,000 |
| [124] | RQC-III | 256 bits | Intel Core i7-7820X @ 3.6 GHz, 16 GB of RAM | 1,820,000 | 3,550,000 | 23,200,000 |
| [125] | SABER KEM (LightSABER) | 125-169 bits | Intel Core i5-7200U @ 2.50 GHz | 85,474 | 108,927 | 119,868 |
| [125] | SABER KEM | 203-244 bits | Intel Core i5-7200U @ 2.50 GHz | 163,333 | 196,705 | 215,733 |
| [125] | SABER KEM (FireSABER) | 283-338 bits | Intel Core i5-7200U @ 2.50 GHz | 259,504 | 308,277 | 341,654 |
| [126] | SIKE (SIKEp434) | 128 bits | Intel Core i7-6700 @ 3.4 GHz | 1,047,991,000 | 1,482,681,000 | 1,790,304,000 |
| [127] | Three Bears (BabyBear CCA) | 154-190 bits | Intel Core i3-6100U @ 2.3 GHz | 41,000 | 60,000 | 101,000 |
| [127] | Three Bears (MamaBear CCA) | 235-241 bits | Intel Core i3-6100U @ 2.3 GHz | 79,000 | 96,000 | 156,000 |
| [127] | Three Bears (PapaBear CCA) | 314-317 bits | Intel Core i3-6100U @ 2.3 GHz | 118,000 | 145,000 | 211,000 |

TABLE IX
CHARACTERISTICS OF THE MICROPROCESSORS USED FOR THE EVALUATION OF POST-QUANTUM ALGORITHMS.

| References | Microprocessor | Clock Frequency | Class | Core | Typical TDP | Release Data | Passmark Rank | Passmark Average Mark | Passmark Single Thread Rating |
|---|---|---|---|---|---|---|---|---|---|
| [111] | Intel Core i5-6260U | 1.8 GHz | Laptop | Skylake | 15 W | Q4 2015 | 982 | 4,362 | 1593 |
| [112] | Intel Xeon E3-1220 v3 | 3.1 GHz | Server | Haswell | 80 W | Q1 2011 | 596 | 7,163 | 1945 |
| [113] | Intel Core i7-4770K | 3.5 GHz | Desktop | Haswell | 84 W | Q2 2013 | 318 | 10,075 | 2250 |
| [114], [126] | Intel Core i7-6700 | 3.4 GHz | Desktop | Skylake | 65 W | Q2 2015 | 323 | 10,003 | 2154 |
| [115], [122], [124] | Intel Core i7-7820X | 3.6 GHz | Desktop | Skylake | 140 W | Q2 2017 | 81 | 18,489 | 2401 |
| [116] | Intel Core i7-4770S | 3.1 GHz | Desktop | Haswell | 65 W | Q2 2013 | 383 | 9,348 | 2174 |
| [117] | Intel Core i5-6600 | 3.3 GHz | Desktop | Skylake | 65 W | Q2 2015 | 536 | 7,778 | 2095 |
| [118] | Intel Core i7-4770K | 3.5 GHz | Desktop | Haswell | 84 W | Q2 2013 | 318 | 10,075 | 2250 |
| [120] | Intel Xeon E3-1275 v3 | 3.5 GHz | Server | Haswell | 95 W | Q2 2013 | 333 | 9,915 | 2214 |
| [121] | Intel Xeon E5-2667 v2 | 3.3 GHz | Server | Sandy Bridge | 130 W | Q1 2014 | 83 | 22,568 | 2023 |
| [125] | Intel Core i5-7200U | 2.5 GHz | Laptop | Kaby Lake | 15 W | Q4 2016 | 932 | 4,602 | 1722 |
| [127] | Intel Core i3-6100U | 2.3 GHz | Laptop | Skylake | 15 W | Q4 2015 | 1,143 | 3,603 | 1302 |

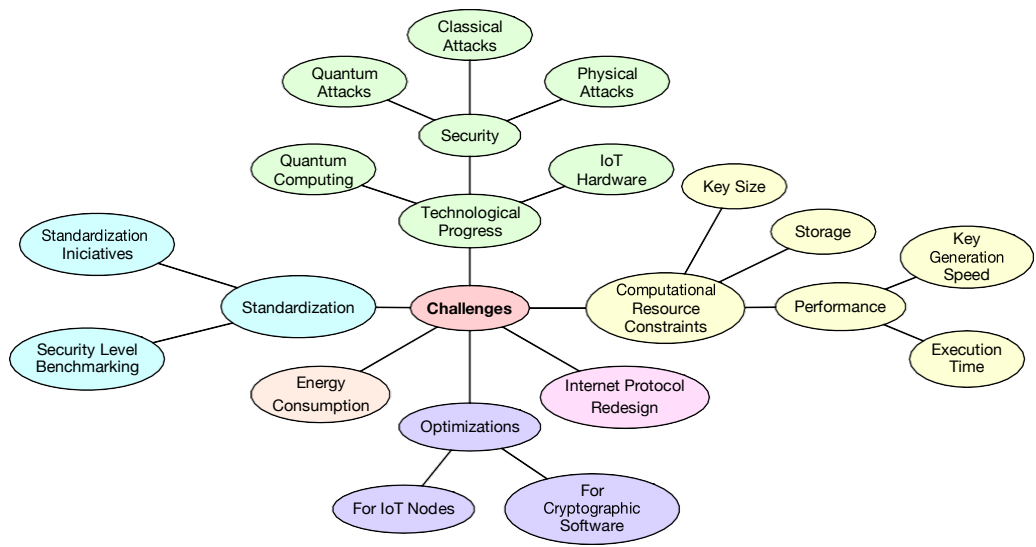

Fig. 4. Main challenges in the development of post-quantum cryptosystems for IoT systems.

Grover's algorithm, this seems to be too conservative, since the required quantum hardware will be far more expensive than the one used by classical computers [14].

*7) Fast IoT hardware evolution:* The selection of the IoT platforms to be evaluated by today's IoT developers is not straightforward, since what are currently considered low-end devices are obviously less powerful than the devices that will exist in 20 years. Therefore, to avoid choosing hardware that is not representative, at least three different groups of IoT devices will have to be distinguished: near-term IoT devices, middle-term devices (i.e., devices that would be considered low-end devices in the middle term) and long-term devices (i.e., devices that would be considered low-end devices by the time large-scale quantum computers will be available for breaking public-key algorithms). For such devices, energy-efficient post-quantum cryptographic systems will have to be developed in order to maintain a trade-off between performance, computational resources and energy consumption.

*8) Need for IoT-node specific optimizations:* Optimizations will have to be designed and performed on post-quantum algorithms when adapting them to IoT devices. For instance, post-quantum lattice-based cryptosystems will need to accelerate recurrent lattice operations (e.g., speeding-up polynomial multiplications) and to minimize consumed energy and running time (and, as a consequence, to reduce key size). Similarly, protocols like SIDH will need to perform efficient double-point multiplications, to accelerate isogeny computation or to speed up modular arithmetic algorithms.

Assembler code for IoT microcontrollers will also have to be optimized depending on the architecture of the selected embedded devices, focusing on the acceleration of integer arithmetic and following traditional assembler optimizations like loop optimization techniques (e.g., loop unrolling, loop pipelining), instruction reordering or register optimized allocation.

*9) Need for cryptographic software optimizations:* Optimizations are also needed on cryptographic software that runs on other devices of an IoT network, like servers, desktop/laptop computers or smartphones. Such optimizations should be aimed at increasing speed (i.e., at minimizing the CPU cycles required by cryptographic operations) and, ideally, to minimize energy consumption in battery-dependent devices.

*10) IoT node hardware unsuitability:* Certain theoretically designed post-quantum cryptosystems may be unsuitable for some IoT devices due to the demanded computational resources or to the consumed energy. To prevent these issues, strict computational and power consumption requirements will have to be established during the theoretical design phase so that, in the worst case, at least one of the three groups of IoT devices (i.e., short-term, middle-term and long-term low-power IoT devices) will be able to run the proposed algorithms appropriately.

*11) Enhanced IoT device physical security:* In order to test and demonstrate the robustness of the developed and optimized post-quantum cryptosystems, it is also necessary to evaluate their physical security, which is essential in an IoT context. It is important to note that, although the proposed post-quantum systems will have to withstand mathematical attacks (even when they are performed by quantum computers), the implementation of the proposed algorithms may be vulnerable to physical attacks, since an attacker may have physical access to IoT devices that execute the implemented algorithms. Therefore, the proposed cryptosystems will have to be designed and evaluated with the objective of avoiding the main physical attacks:

- *ı* Timing attacks. They consist in exploiting the relationship between the execution time and the processed secret data (or the operations performed on such data).
- *ı* Power analysis attacks. Simple Power Analysis attacks are similar to timing attacks, but they exploit the information on the device consumed power when certain operations are executed. In the case of Differential Power Analysis, the power is associated with the data managed by the device.
- *ı* Fault attacks. They are able to induce a fault in the circuitry of an IoT device (e.g., by altering the power supply voltage level, by inducing strong electric/magnetic fields, by overclocking the device clock or by applying high temperatures on the device) in order to obtain information on the private key.

To avoid the aforementioned attacks, different countermeasures will have to be proposed, like the addition of redundant noise to equalize power consumption, private-key splitting (the key is split into shares so that it can only be recovered by assembling all the shares or a minimum subset of them) or the development of constant-time implementations. In addition, the power consumption of the proposed countermeasures will have to be quantified with the objective of achieving the best trade-off between security and energy efficiency.

### B. Future trends

Successful post-quantum IoT cryptosystems can lead to enhance the security of a number of relevant fields whose applications rely heavily on resource-constrained and battery-dependent IoT devices, like home automation [202], [203], [204], smart cities [205], [206], [207], [208], precision agriculture [209], smart garments [210] or Industry 4.0 smart factories [211], [212], [213]. Moreover, the obtained results may benefit the creation and improvement of post-quantum IoT networking architectures that make use of resource-constrained devices [164]. Furthermore, new energy-efficient post-quantum IoT solutions can have a remarkable impact on fields and applications where medium and long-term security are essential, like Defense and Public Safety [214], mission-critical scenarios [215] or smart healthcare [216].

In order to create efficient post-quantum IoT systems, the following contributions would be ideally needed:

- *ı* Novel methodologies will have to address the analysis, quantification and mitigation of quantum computing impact on the different theoretical components of an IoT network.
- *ı* New theoretical insights are still needed on the analysis, design and implementation of resource-constrained post-quantum devices.
- *ı* Methodologies will also be needed for testing and validating energy-efficient quantum-resistant algorithms for resource-constrained IoT devices.
- *ı* Energy-efficient post-quantum IoT architectures will have to be designed and implemented for small-scale scenarios (e.g., home automation), medium-scale environments (e.g., a smart campus [217], a smart factory) or large-scale locations (e.g., a smart city).
- *ı* Post-quantum energy-efficient internet protocols specific for resource-constrained devices will have to be designed and implemented.
- *ı* New coding schemes and techniques will have to be designed and optimized for developing post-quantum IoT code-based cryptosystems.
- *ı* It will be necessary to model mathematically post-quantum IoT node/architecture performance and power consumption.
- *ı* Although quantum-resistant key establishment has been traditionally performed through classical computational methods, researchers will have to study deeper the key establishment physics-based methods that are collectively known as Quantum-Key Distribution (QKD) [46].

### VIII. Conclusion

Although in the last years the interest in post-quantum cryptography has grown significantly, there is still much research to be carried out in its application to IoT. This article provided a thorough survey on the multiple aspects that influence post-quantum IoT development: the need for evolving from pre-quantum security, the most relevant projects and standardization initiatives, the latest post-quantum public-key cryptosystems for potential IoT applications and the main challenges and trends that will arise in the near and middle future. As a result of such contribution, this article provides useful guidelines for the next generation of IoT developers that seek to create quantum-resistant solutions.

### References

[1] HIS, "Internet of Things (IoT) Connected Devices Installed Base Worldwide from 2015 to 2025 (In Billions)". Available online: https://www.statista.com/statistics/471264/ iot-number-of-connected-devices-worldwide/ (accessed on October 2019).

[2] R. L. Rivest, A. Shamir, and L. M. Adleman, "A method for obtaining digital signatures and public-key cryptosystems," *Communications of the ACM*, vol. 21, no. 2, pp. 120-126, Feb. 1978.

[3] W. Diffie and M. E. Hellman, "New directions in cryptography,", *IEEE Transactions on Information Theory*, vol. 22, no. 6, pp. 644-654, Nov. 1976.

[4] IETF, "RFC 8446: The Transport Layer Security (TLS) Protocol Version 1.3", Aug. 2018.

[5] IETF, "RFC 3156: MIME security with OpenPGP", Aug. 2001.

[6] VISA, "Message Level Encryption Documentation". Available online:

[7] R. R. Jueneman, "Electronic document authentication," *IEEE Network*, vol. 1, no. 2, pp. 17-23, Apr. 1987.

[8] C. K. Williams, "Configuring Enterprise Public Key Infrastructures to Permit Integrated Deployment of Signature, Encryption and Access Control Systems". In *Proceedings of IEEE Military Communications Conference*, Atlantic City, USA, Oct. 2005.

[9] C. H. Tseng, S.-H. Wang and W.-J. Tsaur, "Hierarchical and Dynamic Elliptic Curve Cryptosystem Based Self-Certified Public Key Scheme for Medical Data Protection,", *IEEE Transactions on Reliability*, vol. 64, no. 3, pp. 1078-1085, Sep. 2015.

[10] N. Koblitz, "Elliptic curve cryptosystems," *Mathematics of Computation*, vol. 48, no. 177, pp. 203-209, Jan. 1987.

[11] V. S. Miller, "Use of elliptic curves in cryptography,". In *Proceedings of Advances in Cryptology*, LNCS 218, pp. 417-426, Aug. 1985.

[12] T. Kleinjung, K. Aoki, J. Franke, A. K. Lenstra, E. Thomé, J. W. Bos, P. Gaudry, A. Kruppa, P. L. Montgomery, D. A. Osvik, H. te Riele, A. Timofeev and P. Zimmermann, "Factorization of a 768-Bit RSA Modulus". In *Proceedings of the 30th annual conference on Advances in cryptology*, Santa Barbara, USA, Aug. 2010.

[13] A. Pellegrini, V. Bertacoa and T. Austin, "Fault-based attack of RSA authentication". In *Proceedings of the Conference on Design, Automation and Test in Europe*, Dresden, Germany, Mar. 2010.

[14] National Institute of Standards and Technology (NIST), NIST- IR-8105 (draft): "Report on Post-Quantum Cryptography", Apr. 2016.

[15] Committee on National Security Systems (NSS), "CNSS Advisory Memorandum Information Assurance 02-15: Use of Public Standards for the Secure Sharing of Information among National Security Systems", July 2015.

[16] N. Koblitz and A. Menezes, "A Riddle Wrapped in an Enigma," *IEEE Security & Privacy*, vol. 14, no. 6, pp. 34-42, Dec. 2016.

[17] J. Cartwright, "NSA keys into quantum computing," *Physics World*, vol. 27, no. 2, Feb. 2014.

[18] M. Mosca, "Cybersecurity in an Era with Quantum Computers: Will We Be Ready?," *IEEE Security & Privacy*, vol. 16, no. 5, pp. 38-41, Sep./Oct. 2018.

[19] M. Baldi, P. Santini and G. Cancellieri, "Post-quantum cryptography based on codes: State of the art and open challenges". In *Proceedings of the AEIT International Annual Conference*, Cagliari, Italy, Sep. 2017.

[20] C. Cheng, R. Lu, A. Petzoldt and T. Takagi, "Securing the Internet of Things in a Quantum World," *IEEE Communications Magazine*, vol. 55, no. 2, pp. 116-120, Feb. 2017.

[21] S. K. Routray, M. K. Jha, L. Sharma, R. Nyamangoudar, A. Javali and S. Sarkar, "Quantum cryptography for IoT: APerspective". In *Proceedings of the International Conference on IoT and Application (ICIOT)*, Nagapattinam, India, pp. 1-4, May 2017.

[22] R. Chaudhary, G. S. Aujla, N. Kumar and S. Zeadally, "Lattice-Based Public Key Cryptosystem for Internet of Things Environment: Challenges and Solutions," *IEEE Internet of Things Journal*, vol. no. 3, pp. 4897-4909, June 2019.

[23] P. Fraga-Lamas and T. M. Fernández-Caramés, "Reverse engineering the communications protocol of an RFID public transportation card". In *Proceedings of the 2017 IEEE International Conference on RFID*, Phoenix, United States, May 2017.

[24] T. M. Fernández-Caramés, P. Fraga-Lamas, M. Suárez-Albela and L. Castedo, "A methodology for evaluating security in commercial RFID systems". In *Radio Frequency Identification*, IntechOpen, Nov. 2017.

[25] T. M. Fernández-Caramés, P. Fraga-Lamas, M. Suárez-Albela and M., L. Castedo, "Reverse Engineering and Security Evaluation of Commercial Tags for RFID-Based IoT Applications," *Sensors*, vol. 2016, no. 17, Dec. 2016.

[26] C. H. Bennett, E. Bernstein, G. Brassard and U. Vazirani, "Strengths and weaknesses of quantum computing," *SIAM Journal on Computing*, vol. 26, no. 5, Oct. 1997.

[27] G. Brassard, P. Høyer and A. Tapp, "Quantum Cryptanalysis of Hash and Claw-Free Functions". In *Proceedings of the Latin American Symposium on Theoretical Informatics*, Valdivia, Chile, Mar. 2006.

[28] L. K. Grover, "A fast quantum mechanical algorithm for database search". In *Proceedings of the 28th Annual ACM Symposium on the Theory of Computing*, Philadelphia, USA, May 1996.

[29] National Institute of Standards and Technology (NIST), "FIPS 186-2: Digital Signature Standard (DSS)", Jan. 2000.

[30] P. Shor, "Polynomial-Time Algorithms for Prime Factorization and Discrete Logarithms on a Quantum Computer," *SIAM Journal on Computing*, vol. 26, no. 5, pp. 1484-1509, Oct. 1997.

[31] J. Proos and C. Zalka, "Shor's Discrete Logarithm Quantum Algorithm for Elliptic Curves," *Quantum Info. Comput.*, vol. 3, no. 4, pp. 317-344, July 2003.

[32] E. Barker (NIST), "Recommendation for Key Management - Part 1: General (Revision 4)", Jan. 2016.

[33] Y. Gao, X. Chen, Y. Chen, Y. Sun, X. Niu and Y. Yang, "A Secure Cryptocurrency Scheme Based on Post-Quantum Blockchain," *IEEE Access*, vol. 6, pp. 27205-27213, Apr. 2018.

[34] P. Zhang, H. Jiang, Z. Zheng, P. Hu and Q. Xu, "A New Post-Quantum Blind Signature From Lattice Assumptions," *IEEE Access*, vol. 6, pp. 27251-27258, May 2018.

[35] Y. Qassim, M. E. Magaña and A. Yavuz, "Post-quantum hybrid security mechanism for MIMO systems". In *Proceedings of the International Conference on Computing, Networking and Communications*, Santa Clara, USA, pp. 684-689, Jan. 2017.

[36] V. Clupek, L. Malina and V. Zeman, "Secure digital archiving in post-quantum era". In *Proceedings of the 38th International Conference on Telecommunications and Signal Processing*, Prague, Czech Republic, pp. 622-626, July 2015.

[37] PQCRYPTO project official web page. Available online: https: pqcrypto.eu.org (accessed on October 2019)

[38] SAFEcrypto project official web page. Available online: https://www. safecrypto.eu (accessed on October 2019)

[39] CryptoMathCREST project official web page. Available online: https: //cryptomath-crest.jp/english (accessed on October 2019)

[40] PROMETHEUS official web page. Available online: http://prometheuscrypt.gforge.inria.fr (accessed on October 2019)

[41] PQCrypto contributions web page. Available online: https://cordis. europa.eu/project/rcn/194347/results (accessed on October 2019)

[42] SAFEcrypto contributions web page. Available online: https://cordis. europa.eu/project/rcn/194240/results/en (accessed on October 2019)

[43] SAFEcrypto software library on GitHub. Available online: https: //github.com/safecrypto/libsafecrypto (accessed on October 2019)

[44] PROMETHEUS publications web page. Available online: http:// prometheuscrypt.gforge.inria.fr/articles.htm(accessed on October 2019)

[45] ETSI, White Paper: "Quantum Safe Cryptography and Security; An introduction, benefits, enablers and challenges", June 2015.

[46] ETSI, White Paper: "Implementation Security of Quantum Cryptography; Introduction, challenges, solutions", July 2018.

[47] ETSI/IQC 2018 Quantum Safe Workshop web page. Available online: https://www.etsi.org/news-events/events/1296-etsi-iqc-quantum-safe-workshop-2018 (accessed on October 2019)

[48] ETSI's quantum-safe security working group web page. Available online: https://portal.etsi.org/tbsitemap/cyber/cyberqsctor.aspx (accessed on October 2019)

[49] ETSI Technical Committee Cyber Working Group on Quantum-Safe Cryptography, Available online: https://portal.etsi.org/TBSiteMap/ CYBER/CYBERQSCToR.aspx (accessed on October 2019)

[50] ETSI Technical Committee Cyber Working Group on Quantum-Safe Cryptography, "Quantum-Safe Cryptography (QSC); Limits to Quantum Computing applied to symmetric key sizes". Available online: https://www.etsi.org/deliver/etsi gr/QSC/001 099/006/01.01.01 _ 60/gr_QSC006v010101p.pdf (accessed on October 2019)

[51] ETSI Technical Committee Cyber Working Group on Quantum-Safe Cryptography, "Quantum-Safe Cryptography; Quantum-Safe threat assessment". Available online: https://www.etsi.org/deliver/etsi gr/QSC/ 001_099/004/01.01.01_60/gr_QSC004v010101p.pdf

[52] NIST's Workshop on cybersecurity in a post-quantum world web page. Available online: https://www.nist.gov/news-events/events/2015/ 04/workshop-cybersecurity-post-quantum-world (accessed on October 2019)

[53] NIST's announcement of the First Post-Quantum Cryptography Standardization Conference. Available online: https://csrc.nist.gov/events/ 2018/first-pqc-standardization-conference (accessed on October 2019)

[54] Announcement of the NIST's Call for Proposals for public-key post-quantum cryptography algorithms. Available online: https://bit.ly/ 2hKONFb (accessed on October 2019)

[55] NIST's second round announcement on the call for proposals of post-quantum cryptosystems. Available online: https://csrc.nist.gov/news/2019/pqc-standardization-process-2nd-round-candidates (accessed on October 2019)

[56] Crypto Forum Research Group official web page. Available online: https://irtf.org/cfrg (accessed on October 2019)

[57] IETF, "Quantum-Safe Hybrid (QSH) Ciphersuite for Transport Layer Security (TLS) version 1.3", Oct. 2016.

[58] IETF Internet-Draft on post-quantum IKEv2. Available online: https://tools.ietf.org/html/draft-ietf-ipsecme-qr-ikev2-07 (accessed on October 2019)

[59] IETF Internet-Draft on the transition from classical to post-quantum cryptography. Available online: https://datatracker.ietf.org/doc/draft-hoffman-c2pq (accessed on October 2019)

[60] IETF, "RFC 8391 - XMSS: eXtended Merkle Signature Scheme". Available online: https://datatracker.ietf.org/doc/rfc8391/ (accessed on October 2019)

[61] IETF, "RFC 8554 - Leighton-Micali Hash-Based Signatures". Available online: https://datatracker.ietf.org/doc/rfc8554/ (accessed on October 2019)

[62] ISO/IEC JTC 1/SC 27 (Working group on IT Security techniques) web page. Available online: https://www.iso.org/committee/45306.html (accessed on October 2019).

[63] IEEE, "1363.1-2008 - IEEE Standard Specification for Public Key Cryptographic Techniques Based on Hard Problems over Lattices", Mar. 2009.

[64] ANSI ASC X9, White Paper: "ASC X9 IR01-2019 - Quantum Computing Risks to the Financial Services Industry", Feb. 2019.

[65] ANSI ASC X9, "ANSI X9.98-2010 (R2017): Lattice-Based Polynomial Public Key Establishment Algorithm for the Financial Services Industry", Feb. 2017.

[66] PQCrypto deliverable on standardization initiatives. Available online: https://pqcrypto.eu.org/deliverables/d5.2-final.pdf (accessed on October 2019)

[67] R. J. McEliece, "A Public-Key Cryptosystem Based On Algebraic Coding Theory", Deep Space Network Progress Report, DSN PR 42-44, pp. 114-116, Jan.-Feb. 1978.

[68] E. R. Berlekamp, "Goppa Codes," IEEE Transactions on information theory, vol. IT-19, No. 5, Sep. 1973.

[69] E. R. Berlekamp, R. J. McEliece and H. C. A. van Tilborg, "On the inherent intractability of certain coding problems," IEEE Trans. Information Theory, vol. 24, no. 3, pp. 384-386, May 1978.

[70] W. Lee, J.-S. No and Y.-S. Kim, "Punctured Reed-Muller code-based McEliece cryptosystems," IET Communications, vol. 11, no. 10, pp. 1543-1548, July 2017.

[71] H. Niederreiter, "Knapsack-type cryptosystems and algebraic coding theory", Problems of Control and Information Theory, vol. 15, pp. 159-166, Jan. 1986.

[72] N. T. Courtois, M. Finiasz and N. Sendrier, "How to Achieve a McEliece-Based Digital Signature Scheme," in Proceedings of the International Conference on the Theory and Application of Cryptology and Information Security, Gold Coast, Australia, Dec. 2001.

[73] M. Abdalla, J. H. An, M. Bellare and C. Namprempre, "From Identification to Signatures via the Fiat-Shamir Transform: Minimizing Assumptions for Security and Forward-Security", in Proceedings of the International Conference on the Theory and Applications of Cryptographic Techniques, Amsterdam, The Netherlands, Apr.-May 2002.

[74] S. M. E. Y. Alaoui, P. L. Cayrel, R. El Bansarkhani and G. Hoffmann, "Code-based identification and signature schemes in software," in Proceedings of the International Conference on Availability, Reliability and Security, Regensburg, Germany, Sep. 2013.

[75] D. J. Bernstein, J. Buchman and E. Dahmen, "Post-Quantum Cryptography", Springer, 2009.

[76] J. Ding, A. Petzoldt and L.-C. Wang, "The cubic Simple Matrix Encryption Scheme". In Proceedings of PQCrypto, Waterloo, Canada, Oct. 2014.

[77] J. Ding, "A new Variant of the Matsumoto-Imai Cryptosystem through Perturbation". In Proceedings of the International Workshop on Public Key Cryptography, Singapore, Singapore, Mar. 2004.

[78] J. Ding and D. Schmidt, "Cryptanalysis of HFEv and Internal Perturbation of HFE". In Proceedings of the International Workshop on Public Key Cryptography, Les Diablerets, Switzerland, Jan. 2005.

[79] J. Patarin, "Hidden Field equations (HFE) and Isomorphisms of Polynomials (IP)". In Proceedings of EUROCRYPT, Saragossa, Spain, May 1996.

[80] N. T. Courtois, "On Multivariate Signature-Only Public Key Cryptosystems", IACR Cryptology ePrint Archive, Apr. 2001.

[81] Chen, J.-M. and Yang, B.-Y., "Tame Transformation Signatures with Topsy-Turvy Hashes". In Proceedings of IWAP, Taipei, Taiwan, Oct. 2002.

[82] Wang, L.-C., Hu, Y.-H., Lai, F., Chou, C.-Y. and Yang, B.-Y., "Tractable Rational Map Signature". In Proceedings of the International Workshop on Public Key Cryptography, Les Diablerets, Switzerland, 23-36 Jan. 2005.

[83] J. Ding, B.-Y. Yang,C.-H. O. Chen, M.-S. Chen and C. M. Cheng, "New Differential-Algebraic Attacks and Reparameterization of Rainbow". In Proceedings of the International Conference on Applied Cryptography and Network Security, New York, USA, June 2008.

[84] J. Blömer, S. Naewe, "Sampling Methods for Shortest Vectors, Closest Vectors and Successive Minima." In Proceedings of the International Colloquium on Automata, Languages, and Programming, Wroclaw, Poland, July 2007.

[85] J. Hoffstein, J. Pipher, J. H. Silverman, "NTRU: A ring-based public key cryptosystem". In Proceedings of the Third International Symposium on Algorithmic Number Theory, Portland, United States, June 1998.

[86] D. Stehlé and R. Steinfeld, "Making NTRU as Secure as Worst-Case Problems over Ideal Lattices". In Proceedings of the Annual International Conference on the Theory and Applications of Cryptographic Techniques, Tallinn, Estonia, May 2011.

[87] R. Chaudhary, A. Jindal, G. S. Aujla, N. Kumar, A. K. Das and N. Saxena, "LSCSH: Lattice Based Secure Cryptosystem for Smart Healthcare in Smart Cities Environment," IEEE Communications Magazine, vol. 56, no. 4, pp. 24-32, Apr. 2018.

[88] R. Chaudhary, G. S. Aujla, N. Kumar, A. K. Das, N. Saxena and J. J. P. C. Rodrigues, "LaCSys: Lattice Based Cryptosystem for Secure Communication in Smart Grid Environment,". In Proceedings of the IEEE International Conference on Communications, Kansas City, United States, May 2018.

[89] G. S. Aujla, R. Chaudhary, K. Kaur, S. Garg, N. Kumar and R. Ranjan, "SAFE: SDN-Assisted Framework for Edge–Cloud Interplay in Secure Healthcare Ecosystem," IEEE Transactions on Industrial Informatics, vol. 15, no. 1, pp. 469-480, Jan. 2019.

[90] R. Lindner and C. Peikert, "Better key sizes (and attacks) for LWE-based encryption". In Proceedings of the Cryptographers' Track at the RSA Conference, San Francisco, California, Feb. 2011.

[91] V. Lyubashevsky, C. Peikert, O. Regev, "A Toolkit for Ring-LWE Cryptography". In Proceedings of EUROCRYPT 2013, Athens, Greece, May 2013.

[92] E. Alkim, L. Ducas, T. Pöppelmann, P. Schwabe, "Post-quantum key exchange - A new hope". In Proceedings of the USENIX Security Symposium, pp. 327-343, Aug. 2016.

[93] M. Ajtai, "Generating hard instances of lattice problems". In Proceedings of the twenty-eighth annual ACM symposium on Theory of Computing, Philadelphia, United States, May 1996.

[94] V. Lyubashevsky, "Lattice signatures without trapdoors". In Proceedings of the Annual International Conference on the Theory and Applications of Cryptographic Techniques, Cambridge, UK, Apr. 2012.

[95] L. Ducas, A. Durmus, T. Lepoint and V. Lyubashevsky, "Lattice Signatures and Bimodal Gaussians", in Proceedings of the Annual Cryptology Conference, Santa Barbara, United States, Aug. 2013.

[96] T. Pöppelmann, L. Ducas and T. Güneysu, "Enhanced Lattice-Based Signatures on Reconfigurable Hardware", in Proceedings of IACR-CHES, Busan, Korea, Apr. 2014.

[97] J. Ding, X. Xie and X. A. Lin, "A simple provably secure key exchange scheme based on the learning with errors problem," in IACR Cryptology ePrint Archive, Report 2012/688, Dec. 2012.

[98] C. Peikert, "Lattice cryptography for the Internet," Lecture Notes in Computer Science, vol. 8772, pp. 197-219, July 2014.

[99] J. W. Bos, C. Costello, M. Naehrig, D. Stebila, "Post-quantum key exchange for the TLS protocol from the ring learning with errors problem," in Proceedings of the IEEE Symposium on Security and Privacy, San José, United States, May 2015.

[100] A. Rostovtsev and A. Stolbunov, "Public-key cryptosystem based on isogenies", Cryptology ePrint Archive, Report 2006/145, 2006.

[101] A. M. Childs, D. Jao and V. Soukharev, "Constructing elliptic curve isogenies in quantum subexponential time," Journal of Mathematical Cryptology, vol. 8, no. 1, Oct. 2013.

[102] J.-F. Biasse, D. Jao, A. Sankar, "A quantum algorithm for computing isogenies between supersingular elliptic curves," Lecture Notes in Computer Science, vol. 8885, pp. 428-442, Oct. 2014.

[103] L. De Feo, D. Jao, J. Plût, "Towards quantum-resistant cryptosystems from supersingular elliptic curve isogenies", Journal of Mathematical Cryptology, vol. 8, no. 3, pp. 209-247, Jun. 2014.

[104] C. Costello, P. Longa and M. Naehrig, "Efficient algorithms for super-singular isogeny Diffie-Hellman," *Cryptology ePrint Archive*, Report 2016/413, 2016.

[105] X. Sun, H. Tian and Y. Wang, "Toward Quantum-Resistant Strong Designated Verifier Signature from Isogenies". In *Proceedings of the Fourth International Conference on Intelligent Networking and Collaborative Systems*, Bucharest, Romania, Sep. 2012.

[106] C. Costello, D. Jao, P. Longa, M. Naehrig, J. Renes and D. Urbanik, "Efficient Compression of SIDH Public Keys". In *Proceedings of the Annual International Conference on the Theory and Applications of Cryptographic Techniques*, Paris, France, Apr. 2017.

[107] R. Azarderakhsh, D. Jao, K. Kalach, B. Koziel and C. Leonardi, "Key compression for isogeny-based cryptosystems". In *Proceedings of the 3rd ACM International Workshop on ASIA Public-Key Cryptography*, Xi'an, China, May-June, 2016.

[108] Google Blog on Google's experiments with a hybrid cryptosystem. Available online: https://security.googleblog.com/2016/07/experimenting-with-post-quantum.html (accessed on October 2019)

[109] Tor project test with hybrid handshake. Available online: https://gitweb.torproject.org/user/isis/torspec.git/plain/proposals/XXX-newhope-hybrid-handshake.txt?h=draft/newhope (accessed on October 2019)

[110] ETSI, Report: "Quantum-Safe Cryptography (QSC); Quantum-safe algorithmic framework", July 2016.

[111] Official web page of BIKE suite. Available online: https://bikesuite.org (accessed on October 2019)

[112] Official web page of Classical McEliece. Available online: https://classic.mceliece.org (accessed on October 2019)

[113] Official web page of Kyber. Available online: https://pq-crystals.org/kyber/index.shtml (accessed on October 2019)

[114] Official web page of FrodoKEM. Available online: https://frodokem.org (accessed on October 2019)

[115] Official web page of HQC. Available online: https://pqc-hqc.org (accessed on October 2019)

[116] LAC's NIST submission package. Available online: https://csrc.nist.gov/CSRC/media/Projects/Post-Quantum-Cryptography/documents/round-1/submissions/LAC.zip (accessed on October 2019)

[117] Official web page of LEADcrypt. Available online: https://www.ledacrypt.org/LEDAcrypt/ (accessed on October 2019)

[118] Official web page of New Hope. Available online: https://newhopecrypto.org (accessed on October 2019)

[119] Official web page of NTRU. Available online: https://www.onboardsecurity.com/nist-post-quantum-crypto-submission (accessed on October 2019)

[120] Official web page of NTRU Prime. Available online: https://ntruprime.cr.yp.to (accessed on October 2019)

[121] Official web page of NTS-KEM. Available online: https://nts-kem.io (accessed on October 2019)

[122] Official web page of ROLLO. Available online: https://pqc-rollo.org (accessed on October 2019)

[123] Official web page of Round5. Available online: https://round5.org (accessed on October 2019)

[124] Official web page of RQC. Available online: https://pqc-rqc.org (accessed on October 2019)

[125] Official web page of SABER. Available online: https://www.esat.kuleuven.be/cosic/pqcrypto/saber/ (accessed on October 2019)

[126] Official web page of SIKE. Available online: https://sike.org (accessed on October 2019)

[127] Official Source Forge repository of Three Bears. Available online: https://sourceforge.net/projects/threebears/ (accessed on October 2019)

[128] J. Gubbi, R. Buyya, S. Marusic and M. Palaniswamim, "Internet of Things (IoT): A vision, architectural elements, and future directions,", *Future Gener. Comput. Syst.*, vol. 29, no. 7, pp. 1645-1660, Sep. 2013.

[129] BIKE's documentation for the second round of the NIST Call. Available online: https://bikesuite.org/files/BIKE.pdf (accessed on October 2019)

[130] Paulo S. L. M. Barreto, Shay Gueron,Tim Güneysu, Rafael Misoczki, Edoardo Persichetti, Nicolas Sendrier, Jean-Pierre Tillich, "CAKE: Code-based Algorithm for Key Encapsulation". In *Proceedings of the 16th IMA International Conference on Cryptography and Coding*, Oxford, United Kingdom, Dec. 2017.

[131] R. Misoczki, J.-P. Tillich, N. Sendrier and P. L. S. M. Barreto "MDPC-McEliece: New McEliece variants from moderate density parity-check codes". In *Proceedings of the IEEE International Symposium on Information Theory*, Istambul, Turkey, July 2013.

[132] Daniele Micciancio "Improving lattice based cryptosystems using the hermite normal form". In *Proceedings of the International Cryptography and Lattices Conference*, Providence, United States, pp. 126-145, March 2001.

[133] P.-L. Cayrel, G. Hoffmann and E. Persichetti, "Effcient implementation of a cca2-secure variant of McEliece using generalized Srivastava codes". In *Proceedings of PKC*, Darmstadt, Germany, May 2012.

[134] J.-C. Deneuville, P.Gaborit and G. Zémor, "Ouroboros: A simple, secure and efficient key exchange protocol based on coding theory". In *Proceedings of PQCRYPTO*, Utrecht, The Netherlands, June 2017.

[135] Classical McEliece documentation for the second round of the NIST Call. Available online: https://classic.mceliece.org/nist/mceliece-20171129.pdf (accessed on October 2019)

[136] D. J. Bernstein, T. Lange and C. Peters, "Attacking and defending the McEliece cryptosystem". In *Proceedings of PQCRYPTO*, Cincinnati, United States, Oct. 2008.

[137] A. Canteaut and N. Sendrier, "Cryptanalysis of the original McEliece cryptosystem". In *Proceedings of ASIACRYPT*, Beijing, China, Oct. 1998.

[138] Kyber documentation for the second round of the NIST Call. Available online: https://pq-crystals.org/kyber/data/kyber-specification-round2.pdf (accessed on October 2019)

[139] J. Bos, L. Ducas, E. Kiltz, T. Lepoint, V.Lyubashevsky, J. M. Schanck, P. Schwabe and D. Stehlée, "CRYSTALS – Kyber: a CCA-secure module-lattice-based KEM". In *Proceedings of the IEEE European Symposium on Security and Privacy*, London, United Kingdom, Apr. 2018.

[140] A. Langlois and D. Stehlé "Worst-case to average-case reductions for module lattices," *Designs, Codes and Cryptography*, vol. 75, no. 3, pp. 565-599, June 2015.

[141] FrodoKEM's documentation for the second round of the NIST Call. Available online: https://frodokem.org/files/ FrodoKEM-specification-20190702.pdf (accessed on October 2019)

[142] J. W. Bos, C. Costello, L. Ducas, I. Mironov, M. Naehrig, V. Niko-laenko, A. Raghunathan and D. Stebila "Frodo: Take off the ring! Practical, quantum-secure key exchange from LWE". In *Proceedings of ACM CCS*, Vienna, Austria, Oct. 2016.

[143] HQC documentation for the second round of the NIST Call. Available online: https://pqc-hqc.org/doc/hqc-specification_2019-04-10.pdf (accessed on October 2019)

[144] C. Aguilar-Melchor, O. Blazy, J.-C. Deneuville, P. Gaborit, G. Zémor, "Efficient encryption from random quasi-cyclic codes," *IEEE Transactions on Information Theory*, vol. 64, no. 5, pp. 3927-3943, Feb. 2018.

[145] LEDACrypt's documentation for the second round of the NIST Call. Available online: https://www.ledacrypt.org/documents/LEDAcrypt_spec_latest.pdf (accessed on October 2019)

[146] H. Jiang, Z. Zhang and Z. Ma, "Tighter security proofs for generic key encapsulation mechanism in the quantum random oracle model". In *Proceedings of PQCrypto*, Chongqing, China, May 2019.

[147] New Hope's documentation for the second round of the NIST Call. Available online: https://newhopecrypto.org/data/NewHope_2019_07_10.pdf (accessed on October 2019)

[148] E. Alkim, L. Ducas, T. Pöppelmann, and P. Schwabe "NewHope without reconciliation," *Cryptology ePrint Archive*, Report 2016/1157, 2016.

[149] J. Hoffstein, J. Pipher, J. M. Schanck, J. H. Silverman, W. Whyte, Z. Zhang "Choosing parameters for ntruencrypt". In *Proceedings of the RSA Conference*, San Francisco, United States, Feb. 2017.

[150] NTRU Prime's documentation for the second round of the NIST Call. Available online: https://ntruprime.cr.yp.to/nist/ntruprime-20190330.pdf (accessed on October 2019)

[151] D. J. Bernstein, C. Chuengsatiansup, T. Lange, C. van Vredendaal, "NTRU Prime: reducing attack surface at low cost". In *Proceedings of SAC*, Ottawa, Canada, Aug. 2017.

[152] NTS-KEM's documentation for the second round of the NIST Call. Available online: https://drive.google.com/file/d/1qPsXhK oXJ88M1ec6pRbvvRKaCMQZfsc/view (accessed on October 2019)

[153] ROLLO's documentation for the second round of the NIST Call. Available online: https://pqc-rollo.org/doc/rollo-specification_2019-04-10.pdf (accessed on October 2019)

[154] Round5's documentation for the second round of the NIST Call. Available online: https://round5.org/Supporting Documentation/Round5 Submission.pdf (accessed on October 2019)

[155] RQC's documentation for the second round of the NIST Call. Available online: https://pqc-rqc.org/doc/rqc-specification_2019-04-10.pdf (accessed on October 2019)

[156] J.-P. D'Anvers, A. K. S. S. Roy and F. Vercauteren "Saber: Module-LWR based key exchange, CPA-secure encryption and CCA-secure KEM". In *Proceedings of Africacrypt*, Marrakesh, Morocco, May 2018.

[157] SIKE's documentation for the second round of the NIST Call. Available online: https://sike.org/files/SIDH-spec.pdf (accessed on October 2019)

[158] N. Kshetri, "Can Blockchain Strengthen the Internet of Things?," *IT Professional*, vol. 19, no. 4, pp. 68-72, Aug. 2017.

[159] F. Bonomi, R. Milito, J. Zhu and S. Addepalli, "Fog Computing and its Role in the Internet of Things". In *Proceedings of the First Edition of the MCC Workshop on Mobile Cloud Computing*, Helsinki, Finland, Aug. 2012.

[160] J. S. Preden, K. Tammemäe, A. Jantsch, M. Leier, A. Riid and E. Calis, "The Benefits of Self-Awareness and Attention in Fog and Mist Computing", *Computer*, vol. 48, no. 7, July 2015.

[161] M. Suárez-Albela, P. Fraga-Lamas and T. M. Fernández-Caramés, "A Practical Evaluation on RSA and ECC-Based Cipher Suites for IoT High-Security Energy-Efficient Fog and Mist Computing Devices," *Sensors*, vol. 18, no. 11, Nov. 2018.

[162] M. Suárez-Albela, P. Fraga-Lamas, L. Castedo and T. M. Fernández-Caramés, "Clock frequency impact on the performance of high-security cryptographic cipher suites for energy-efficient resource-constrained IoT devices," *Sensors*, vol. 19, no. 1, Jan. 2019.

[163] I. Froiz-Míguez, T. M. Fernández-Caramés, P. Fraga-Lamas and L. Castedo, "Design, Implementation and Practical Evaluation of an IoT Home Automation System for Fog Computing Applications Based on MQTT and ZigBee-WiFi Sensor Nodes," *Sensors*, vol. 2018, Aug. 2018.

[164] T. M. Fernández-Caramés, P. Fraga-Lamas, M. Suárez-Albela and M. Vilar-Montesinos, "A Fog Computing and Cloudlet Based Augmented Reality System for the Industry 4.0 Shipyard," *Sensors*, vol. 18, June 2018.

[165] T. M. Fernández-Caramés, P. Fraga-Lamas, Suárez-Albela and M. A. D´iaz-Bouza, "A Fog Computing Based Cyber-Physical System for the Automation of Pipe-Related Tasks in the Industry 4.0 Shipyard," *Sensors*, vol. 18, June 2018.

[166] M. Suárez-Albela, T. M. Fernández-Caramés, P. Fraga-Lamas and L. Castedo, "A Practical Evaluation of a High-Security Energy-Efficient Gateway for IoT Fog Computing Applications", *Sensors*, vol. 17, Aug. 2017.

[167] K. Dolui and S. K. Datta, "Comparison of edge computing implementations: Fog computing, cloudlet and mobile edge computing". In *Proceedings of the Global Internet of Things Summit*, Geneva, Switzerland, June 2017. pp. 1-6.

[168] T. M. Fernández-Caramés and P Fraga-Lamas, "A Review on the Use of Blockchain for the Internet of Things," *IEEE Access*, vol. 6, May 2018.

[169] K. Christidis and M. Devetsikiotis, "Blockchains and Smart Contracts for the Internet of Things," *IEEE Access*, vol. 4, pp. 2292-2303, May 2016.

[170] T. M. Fernández-Caramés, I. Froiz-Míguez, O. Blanco-Novoa, P. Fraga-Lamas, "Enabling the Internet of Mobile Crowdsourcing Health Things: A Mobile Fog Computing, Blockchain and IoT Based Continuous Glucose Monitoring System for Diabetes Mellitus Research and Care," *Sensors*, vol. 2019, no. 19, July 2019.

[171] T. M. Fernández-Caramés, O. Blanco-Novoa,I. Froiz-Míguez and P. Fraga-Lamas, "Towards an Autonomous Industry 4.0 Warehouse: A UAV and Blockchain-Based System for Inventory and Traceability Applications in Big Data-Driven Supply Chain Management," *Sensors*, vol. 2019, no. 19, May 2019.

[172] P. Fraga-Lamas and T. M. Fernández-Caramés, "A Review on Blockchain Technologies for an Advanced and Cyber-Resilient Automotive Industry," *IEEE Access*, vol. 2019, no.7, pp. 17578-17598, Jan. 2019.

[173] T. M. Fernández-Caramés and P. Fraga-Lamas, "A Review on the Application of Blockchain for the Next Generation of Cybersecure Industry 4.0 Smart Factories," *IEEE Access*, vol. 2019, no. 7, pp. 45201-45218, Apr. 2019.

[174] O. M. Guillen, T. Pöppelmann, J. M. Bermudo Mera, E. F. Bongenaar, G. Sigl and J. Sepulveda, "Towards post-quantum security for IoT endpoints with NTRU". In *Proceedings of the Design, Automation & Test in Europe Conference & Exhibition*, Lausanne, Switzerland, Mar. 2017.

[175] D. V. Bailey, D. Coffin, A. Elbirt, J. H. Silverman and A. D. Woodbury, "NTRU in Constrained Devices,". In *Proceedings of CHES*, Paris, France, May 2001.

[176] O. Collen Marie, "Efficient NTRU implementation". Master's thesis, Worcester Polytechnic Institute, 2002. Available online: https://www.wpi.edu/Pubs/ETD/Available/etd-0430102-111906/unrestricted/corourke.pdf (accessed on October 2019)

[177] M. Monteverde, "NTRU Software Implementation for Constrained Devices". Master's thesis, Katholieke Universiteit Leuven, Aug. 2008.

[178] A. Boorghany, S. B. Sarmadi and R. Jalili, "On Constrained Implementation of Lattice-Based Cryptographic Primitives and Schemes on Smart Cards," *ACM Trans. Embed. Comput. Syst.*, vol. 14, no. 3, Apr. 2015.

[179] Z. Liu, T. Pöppelmann, T. Oder, H. Seo, S. S. Roy, T. Güneysu, J. Großschädl, H. Kim and I. Verbauwhede, "High-Performance Ideal Lattice-Based Cryptography on 8-bit AVR Microcontrollers," *ACM Transactions on Embedded Computing Systems*, vol. 16, no. 4, Sep. 2017.

[180] Z. Liu, H. Seo, S. S. Roy, J. Großschädl, H. Kim, I. Verbauwhede, "Efficient Ring-LWE encryption on 8-bit AVR processors". In *Proceedings of CHES*, Saint Malo, France, Sep. 2015.

[181] J. Buchmann, F. Göpfert, T. Güneysu, T. Oder and T. Pöppelmann, "High-Performance and Lightweight Lattice-Based Public-Key Encryption". In *Proceedings of the ACM International Workshop on IoT Privacy, Trust, and Security*, Xi'an, China, May 2016.

[182] S. Heyse, I. von Maurich, T. Güneysu,, "Smaller keys for code-based cryptography: QC-MDPC McEliece implementations on embedded devices". In *Proceedings of CHES*, Santa Barbara, United States, Aug. 2013.

[183] T. Eisenbarth, T. Güneysu, S. Heyse and C. Paar, "MicroEliece: McEliece for Embedded Devices". In *Proceedings of CHES*, Lausanne, Switzerland, Sep. 2009.

[184] T. Güneysu and T. Oder, "Towards lightweight Identity-Based Encryption for the post-quantum-secure Internet of Things". In *Proceedings of the 18th International Symposium on Quality Electronic Design*, Santa Clara, United States, Mar. 2017.

[185] E. Alkim, P. Jakubeit and P. Schwabe, "A new hope on ARM Cortex-M". In *Proceedings of the 6th Conference on Security, Privacy, and Applied Cryptography Engineering*, Hyderabad, India, Dec. 2016.

[186] J. Sepulveda, A. Zankl and O. Mischke, "Cache attacks and countermeasures for NTRUEncrypt on MPSoCs: Post-quantum resistance for the IoT". In *Proceedings of the 30th IEEE International System-on-Chip Conference*, Munich, Germany, Sep. 2017.

[187] C. Tsai, M. Hsiao, W. Shen, A. A. Wu and C. Cheng, "A 1.96 mm2 low-latency multi-mode crypto-coprocessor for PKC-based IoT security protocols". In *Proceedings of the IEEE International Symposium on Circuits and Systems*, Lisbon, Portugal, May 2015.

[188] L. Botros, M. J. Kannwischer and P. Schwabe, "Memory-Efficient High-Speed Implementation of Kyber on Cortex-M4". In *Proceedings of AFRICACRYPT 2019*, Rabat, Morocco, July 2019.

[189] T. Pöppelmann, T. Oder and T. Güneysu, "High-Performance Ideal Lattice-Based Cryptography on 8-Bit ATxmega Microcontrollers". In *Proceedings of the 4th International Conference on Cryptology and Information Security in Latin America* Guadalajara, Mexico, Aug. 2015.

[190] S. Heyse, "Implementation of McEliece Based on Quasi-dyadic Goppa Codes for Embedded Devices". In *Proceedings of PQCrypto 2011*, Taipei, Taiwan, Nov.-Dec. 2011.

[191] T. M. Fernández-Caramés, GM. onzález-López and L. Castedo, "FPGA-based vehicular channel emulator for real-time performance evaluation of IEEE 802.11 p transceivers," *EURASIP Journal on Wireless Communications and Networking*, vol. 2010, no. 1, Feb. 2010.

[192] T. M. Fernández-Caramés, GM. onzález-López and L. Castedo, "FPGA-based vehicular channel emulator for evaluation of IEEE 802.11 p transceivers". In *Proceedings of the International Conference on Intelligent Transport Systems Telecommunications*, Lille, France, Oct. 2009.

[193] A. A. Kamal and A. M. Youssef, "An FPGA implementation of the NTRUEncrypt cryptosystem". In *Proceedings of the International Conference on Microelectronics*, Marrakech, Morocc, Dec. 2009.

[194] N. Göttert, T. Feller, M. Schneider, J. Buchman, and S. A. Huss, "On the Design of Hardware Building Blocks for Modern Lattice-Based Encryption Schemes". In *Proceedings of CHES*, Leuven, Belgium, Sep. 2012.

[195] B. Koziel, R. Azarderakhsh, M. Mozaffari Kermani and D. Jao, "Post-Quantum Cryptography on FPGA Based on Isogenies on Elliptic Curves", *IEEE Transactions on Circuits and Systems*, vol. 64, no. 1, Jan. 2017.

[196] T. Pöppelmann and T. Güneysu, "Towards practical lattice-based public- key encryption on reconfigurable hardware". In *Proceedings of*

*the 20th International Conference in Selected Areas in Cryptography*, Burnaby, Canada, Aug. 2013.

[197] S. S. Roy, F. Vercauteren, N. Mentens, D. D. Chen and I. Verbauwhede, "Compact hardware implementation of ring-LWE cryptosystems," *IACR Cryptology ePrint Archive*, vol. 2013, p. 866, Jun. 2014.

[198] T. Pöppelmann and T. Güneysu, "Area optimization of lightweight lattice-based encryption on reconfigurable hardware.". In *Proceedings of the IEEE International Symposium on Circuits and Systems*, Melbourne, Australia, June 2014.

[199] Passmark CPU benchmarks main web page. Available online: https://www.cpubenchmark.net (accessed on October 2019)

[200] A. Shoufan, T. Wink, H. G. Molter, S. A. Huss and E. Kohnert, "A Novel Cryptoprocessor Architecture for the McEliece Public-Key Cryptosystem," *IEEE Trans. Computers*, vol. 59, no. 11, pp. 1533-1546, Nov. 2010.

[201] S. Heyse and T. Güneysu, "Towards One Cycle per Bit Asymmetric Encryption: Code-Based Cryptography on Reconfigurable Hardware". In *Proceedings of CHES*, Leuven, Belgium, Sep. 2012.

[202] M. Suárez-Albela, P. Fraga-Lamas, T. M. Fernández-Caramés, A. Dapena and M. González-López, "Home Automation System Based on Intelligent Transducer Enablers," *Sensors*, vol.16, no. 10, Sep. 2016.

[203] T. M. Fernández-Caramés, "An intelligent power outlet system for the smart home of the Internet of Things," *Int. J. Distrib. Sens. Netw.*, vol. 2015, no. 11, Nov. 2015.

[204] O. Blanco-Novoa, T. M. Fernández-Caramés, P. Fraga-Lamas and L. Castedo, "An Electricity Price-Aware Open-Source Smart Socket for the Internet of Energy," *Sensors*, vol. 2017, no. 3, Mar. 2017.

[205] F. Alam, R. Mehmood, I. Katib, N. N. Albogami and A. Albeshri, "Data Fusion and IoT for Smart Ubiquitous Environments: A Survey," *IEEE Access*, vol. 5, Apr. 2017.

[206] D. L. Hernández-Rojas, T. M. Fernández-Caramés, P. Fraga-Lamas and C. J. Escudero, "A Plug-and-Play Human-Centered Virtual TEDS Architecture for the Web of Things," *Sensors*, vol. 2018, no. 6, June 2018.

[207] D. L. Hernández-Rojas, T. M. Fernández-Caramés, P. Fraga-Lamas and C. J. Escudero, "Design and Practical Evaluation of a Family of Lightweight Protocols for Heterogeneous Sensing through BLE Beacons in IoT Telemetry Applications," *Sensors*, vol. 2018, no. 1, Dec. 2017.

[208] O. Blanco-Novoa, T. M. Fernández-Caramés, P. Fraga-Lamas and L. Castedo, "A Cost-Effective IoT System for Monitoring Indoor Radon Gas Concentration," *Sensors*, vol. 2018, no. 7, July 2018.

[209] J. M. Pérez-Expósito, T. M. Fernández-Caramés, P. Fraga-Lamas and L. Castedo, "VineSens: An Eco-Smart Decision Support Viticulture System," *Sensors*, vol. 17, Feb. 2017.

[210] T. M. Fernández-Caramés and P. Fraga-Lamas, "Towards The Internet of Smart Clothing: A Review on IoT Wearables and Garments for Creating Intelligent Connected E-Textiles," *Electronics*, vol. 2018, no. 7, July 2018.

[211] T. M. Fernández-Caramés and P. Fraga-Lamas, "A Review on Human-Centered IoT-Connected Smart Labels for the Industry 4.0," *IEEE Access*, vol. 6, May 2018.

[212] P. Fraga-Lamas and T. M. Fernández-Caramés, O. Blanco-Novoa and M. A. Vilar-Montesinos, "A Review on Industrial Augmented Reality Systems for the Industry 4.0 Shipyard", *IEEE Access*, vol. 6, Feb. 2018.

[213] P. Fraga-Lamas, D. Noceda-Davila, T. M. Fernández-Caramés, M. A. D´ıaz-Bouza and M. A. Vilar- Montesinos, "Smart Pipe System for a Shipyard 4.0," *Sensors*, vol. 16, Dec. 2016.

[214] P. Fraga-Lamas, T. M. Fernández-Caramés, M. Suárez-Albela, L. Castedo and M. González-López, "A Review on Internet of Things for Defense and Public Safety," *Sensors*, vol. 16, Oct. 2016.

[215] P. Fraga-Lamas, T. M. Fernández-Caramés and L. Castedo, "Towards the Internet of Smart Trains: A Review on Industrial IoT-Connected Railways," *Sensors*, vol. 17, June 2017.

[216] S. M. R. Islam, D. Kwak, M. H. Kabir, M. Hossain and K. Kwak, "The Internet of Things for Health Care: A Comprehensive Survey", *IEEE Access*, vol. 3, Jun. 2015.

[217] P. Fraga-Lamas, M. Celaya-Echarri, P. Lopez-Iturri, L. Castedo, L. Azpilicueta, E. Aguirre, M. Suárez-Albela, F. Falcone and T. M. Fernández-Caramés, "Design and Experimental Validation of a Lo-RaWAN Fog Computing Based Architecture for IoT Enabled Smart Campus Applications," *Sensors*, vol. 2019, no. 19, July 2019.

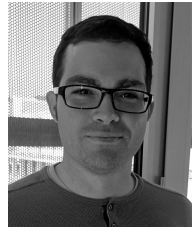

**Tiago M. Fernández-Caramés** (S'08-M'12-SM'15) received his MSc degree and PhD degrees in Computer Science in 2005 and 2011 from University of A Coruña, Spain. Since 2005 he has worked as a researcher and professor for the Department of Computer Engineering of the University of A Coruña inside the Group of Electronic Technology and Communications (GTEC). His current research interests include IIoT/IoT systems, IIoT/IoT security, RFID, wireless sensor networks, Industry 4.0, blockchain and augmented reality.